\documentclass[twocolumn,superscriptaddress,showpacs,aps,amsmath,amssymb]{revtex4}
\usepackage{bm}
\usepackage{graphicx}

\newcommand{\B}{\text{\scriptsize res}}
\newcommand{\s}{\text{\scriptsize sys}}
\newcommand{\SB}{\text{\scriptsize sys-res}}
\newcommand{\T}{{\rm total}}
\newcommand{\dg}{\dagger}

\newcommand{\nl}{\nonumber \\}
\newcommand{\la}{\langle}
\newcommand{\ra}{\rangle}
\newcommand{\up}{\uparrow}
\newcommand{\down}{\downarrow}
\newcommand{\ep}{\epsilon}
\newcommand{\w}{\omega}

\newcommand{\be}{\begin{equation}}
\newcommand{\ee}{\end{equation}}
\newcommand{\bea}{\begin{eqnarray}}
\newcommand{\eea}{\end{eqnarray}}
\newcommand{\bsube}{\begin{subequations}}
\newcommand{\esube}{\end{subequations}}
\newcommand{\Eq}[1]{Eq.\,\eqref{#1}}

\newcommand{\Fig}[1]{Fig.\,\ref{#1}}

\newcommand{\comments}[1]{}

\begin{document}

\title{Hierarchical Liouville-space approach for
  accurate and universal characterization of quantum impurity systems}

\author{ZhenHua~Li}
\affiliation{Department of Physics, Renmin University of China, Beijing
100872,
 China }

\author{NingHua~Tong}
\affiliation{Department of Physics, Renmin University of China, Beijing
100872, China }

\author{Xiao~Zheng}\email{xz58@ustc.edu.cn}
\affiliation{Hefei National Laboratory for Physical Sciences at the
Microscale, University of Science and Technology of China, Hefei, Anhui
230026, China}

\author{Dong~Hou}
\affiliation{Hefei National Laboratory for Physical Sciences at the
Microscale, University of Science and Technology of China, Hefei, Anhui
230026, China}

\author{JianHua~Wei}\email{wjh@ruc.edu.cn}
\affiliation{Department of Physics, Renmin University of China,
Beijing 100872, China }

\author{Jie~Hu}
\affiliation{Department of Chemistry, Hong Kong University of Science
and Technology, Hong Kong, China}
\affiliation{Department of Physics, Capital Normal University, Beijing
100048, China}

\author{YiJing~Yan} \email{yyan@ust.hk}
\affiliation{Hefei National Laboratory for Physical Sciences at the
Microscale, University of Science and Technology of China, Hefei, Anhui
230026, China}
\affiliation{Department of Chemistry, Hong Kong University of Science
and Technology, Hong Kong, China}

\date{Submitted on June~18, 2012; revised on November~23, 2012}

\begin{abstract}

A hierarchical equations of motion (HEOM) based numerical approach is
developed for accurate and efficient evaluation of dynamical
observables of strongly correlated quantum impurity systems.
This approach is capable of describing quantitatively Kondo
resonance and Fermi liquid characteristics, achieving the accuracy
of latest high-level numerical renormalization group approach, as
demonstrated on single-impurity Anderson model systems.
Its application to a two-impurity Anderson model results in
differential conductance versus external bias, which correctly
reproduces the continuous transition from Kondo states of individual
impurity to singlet spin-states formed between two impurities. The
outstanding performance on characterizing both equilibrium and
nonequilibrium properties of quantum impurity systems makes the HEOM
approach potentially useful for addressing strongly correlated lattice
systems in the framework of dynamical mean field theory.

\end{abstract}


\maketitle

Quantum impurity systems cover a broad range of important physical
systems where strong electron-electron (\emph{e-e}) interactions
among a few localized impurities affect crucially the system
properties.
Besides the \emph{e-e} interactions, the impurities are coupled to the
itinerant electrons in surrounding bulk materials, which serve as the
electron reservoir and thermal bath.
Moreover, some extensive strongly correlated systems can be treated
as quantum impurity systems. For instance, the celebrated Hubbard
model can be mapped onto an Anderson impurity system via a
self-consistent dynamical mean-field theory \cite{Geo9613}.
The strong \emph{e-e} interactions give rise to a variety of
intriguing phenomena of prominent many-body nature, such as Kondo
effects, Mott metal-insulator transition, and high-temperature
superconductivity.
Examples of localized impurities are the $d$- or $f$-electrons of
transition metal atoms and electrons trapped in quantum dots.

Accurate characterization of quantum impurity systems is the key to the
understanding of the mechanisms and effects of strong electron
correlations.
This has remained a very challenging task, especially for the
quantitative evaluation of dynamical quantities directly related to
experimental measurements, such as the projected density of states and
spectral function of the localized impurities.
A vast amount of theoretical efforts have been devoted to achieving
this goal, including the quantum Monte Carlo (QMC) approach
\cite{Hir862521, Gul11349}, density matrix renormalization group method
\cite{Whi922863,Vid03147902}, numerical renormalization group (NRG)
method \cite{Wil75773,Bul08395}, many-body perturbation theory
\cite{Kho12075103}, effective/quasi single-particle approaches
\cite{Thy08115333,Kur10236801}, \emph{etc}.
Despite their success in elucidating some fundamental features of
electron correlations, the practicality of existing approaches has been
limited within a few basic models \cite{And6141,Hub63238,Lee8699}.
The reason is mainly twofold: (\emph{i}) the applicability of involving
techniques relies critically on the system configuration, and
(\emph{ii}) the complexity of numerical algorithms increases
dramatically with the number of impurities.
Consequently, generalization of existing approaches \cite{Hir862521,
Gul11349, Whi922863,Vid03147902,Wil75773,Bul08395} to more complex
models is often difficult.
Therefore, an accurate and universal approach capable of addressing
strong correlation effects in general quantum impurity systems is
highly desirable.

In this Letter we propose a general approach based on a hierarchical
equations of motion (HEOM) formalism \cite{Jin08234703} to characterize
quantum impurity systems from the perspective of open dissipative
dynamics. The localized impurities constitute the open system of
primary interest, while the surrounding reservoirs of itinerant
electrons are treated as environment.
The total Hamiltonian 
consists of the interacting impurities ($H_{\s}$), the noninteracting
electron reservoirs ($H_{\B}$), and their couplings
$H_{\SB}=\sum_{\alpha\mu k} (t_{\alpha\mu k}\, \hat{a}^\dag_{\mu}
\hat{d}_{\alpha k} + {\rm H.c.})$.
Here, $\hat{a}_{\mu}^\dag$ and $\hat{a}_{\mu}$ denote the creation and
annihilation operators for impurity state $|\mu\rangle$ (including
spin, space, \emph{etc.}), while $\hat{d}_{\alpha k}^\dag$ and
$\hat{d}_{\alpha k}$ are those for the $\alpha$-reservoir state
$|k\rangle$ of energy $\epsilon_{\alpha k}$. The influence of electron
reservoirs on the impurities is taken into account through the
hybridization functions, $\Delta_{\mu\nu}(\w) \equiv
\sum_{\alpha}\Delta_{\alpha\mu\nu}(\w) = \pi\sum_{\alpha k}
t_{\alpha\mu k}t^\ast_{\alpha\nu k}\, \delta(\w-\ep_{\alpha k})$, in
the absence of applied chemical potentials.

The HEOM that governs the dynamics of open system assumes the form
of \cite{Jin08234703}:
\begin{align}\label{HEOM}
   \dot\rho^{(n)}_{j_1\cdots j_n} =&
   -\Big(i{\cal L} + \sum_{r=1}^n \gamma_{j_r}\Big)\rho^{(n)}_{j_1\cdots j_n}
     -i \sum_{j}\!     
     {\cal A}_{\bar j}\, \rho^{(n+1)}_{j_1\cdots j_nj}
\nl &
    -i \sum_{r=1}^{n}(-)^{n-r}\, {\cal C}_{j_r}\,
     \rho^{(n-1)}_{j_1\cdots j_{r-1}j_{r+1}\cdots j_n}.
\end{align}
The basic variables are the reduced system density operator
$\rho^{(0)}(t) \equiv {\rm tr}_{\B}\,\rho_{\T}(t)$ and auxiliary
density operators, $\{\rho^{(n)}_{j_1\cdots j_n}(t); n=1,\cdots,L\}$,
with $L$ denoting the terminal or truncated tier level. The Liouvillian
of impurities, $\mathcal{L}\,\cdot \equiv \hbar^{-1}[H_{\s}, \cdot\,]$,
may contain both \emph{e-e} interaction and time-dependent external
fields. The superoperators ${\cal A}_{\bar j}$ and ${\cal C}_j$ are
expressed by Eq.\,(S1) of Ref.\,[\onlinecite{Sup1}]. The index $j
\equiv (\sigma\mu m)$ corresponds to the transfer of an electron
to/from ($\sigma=+/-$) the impurity state $\mu$, associated with the
characteristic memory time $\gamma_m^{-1}$.
The total number of distinct $j$-indexes involved is determined by the
preset level of accuracy for decomposing reservoir correlation
functions by exponential functions. Such a number draws the maximum
tier level $L_{\text{max}}$, at which \Eq{HEOM} ultimately terminate
\cite{Sup1}.
The hierarchy is self-contained at $L=2$ for noninteracting $H_{\s}$
\cite{Jin08234703}; while for $H_{\s}$ involving \emph{e-e}
interactions, the solution of \Eq{HEOM} must go through systematic
tests to confirm its convergence versus $L$.
In practice, a relatively low $L$ ($\approx 4$) is usually sufficient
to yield quantitatively converged results for weak and medium
impurity-reservoir couplings.

The details of the HEOM formalism are referred to
Refs.\,[\onlinecite{Jin08234703,Zhe09164708,Zhe121129,Sup1}]. Here,
we focus on some of its key features:
(\emph{i}) It is based on the Feynmann--Vernon path integral
formalism \cite{Fey63118}, with fermionic operators represented by
Grassmann variables \cite{Ryd96}.
(\emph{ii}) It resolves nonperturbatively the combined effects of
impurity-reservoir dissipation, \emph{e-e} interactions, and
non-Markovian memory \cite{Jin08234703}.
(\emph{iii}) The influence of reservoir environment on physical
properties of impurities is taken into account via the hybridization
functions, which enter \Eq{HEOM} through a recently developed optimal
Pad\'{e} spectrum decomposition scheme \cite{Hu10101106, Hu11244106}.
(\emph{iv}) Besides the equilibrium dynamical observables, it is
also capable of addressing nonequilibrium response of quantum
impurity systems to external fields such as laser pulses or applied
voltages \cite{Zhe121129}.

The HEOM approach has been applied to study static and transient
electron transport through quantum dot systems, with which some
interesting phenomena have been revealed, such as the dynamical Coulomb
blockade \cite{Zhe08093016} and dynamical Kondo transition
\cite{Zhe09164708}.

In the framework of HEOM, there are two schemes to evaluate the
dynamical observables of quantum impurity systems.
(\emph{i}) Calculate relevant system correlation/response functions
based on an HEOM-space linear response theory \cite{Sup1}.
The correlation function for two arbitrary system operators $\hat{A}$
and $\hat{B}$ is $\widetilde{C}_{AB}(t) \equiv \la \hat A(t)\hat
B(0)\ra = {\rm tr}_{\T}[\hat A(t)\hat B(0) \rho^{\rm eq}_{\T}(T)]$,
where $\rho^{\rm eq}_{\T}(T)$ is the equilibrium density operator of
the total system. $\widetilde{C}_{AB}(t)$ can be evaluated by using the
quantum Liouville propagator in the HEOM space \cite{Sup1}.
Let $C_{AB}(\w) \equiv \frac{1}{2}\int dt\, e^{i\w t}
\widetilde{C}_{AB}(t)$, which satisfies the detailed balance
relation of $C_{BA}(-\omega)=e^{-\omega/k_B T}C_{AB}(\omega)$. The
corresponding spectral function is
$J_{AB}(\w)  \equiv \frac{1}{2\pi} \int dt\, e^{i \w t}
 \langle \{ \hat A(t), \hat B(0) \} \rangle
 = \frac{1}{\pi} \left(1 + e^{-\omega/k_B T}\right) C_{AB}(\w)$.
In particular, with $\hat A= \hat{B}^\dag = \hat a_{\mu}$, $J_{\hat
a_{\mu}\hat a^{\dg}_{\mu}}(\omega) = A_{\mu}(\omega)$ gives the
spectral density of impurity state $\mu$, which can be measured
experimentally via angle-resolved photoemission spectroscopy
\cite{Dam03473} and scanning tunneling microscope \cite{Kol05085456}.
(\emph{ii}) Solve \Eq{HEOM} for nonequilibrium electronic response
under external perturbation. For instance, the differential conductance
($dI/dV$) can be calculated via the response current under applied
bias, followed by a finite difference analysis. The above two schemes
are completely equivalent for linear response properties.

\begin{figure}
\includegraphics[width=0.9\columnwidth]{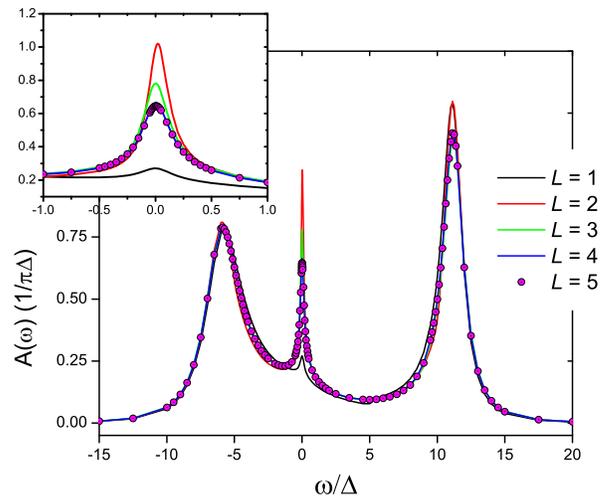}
\caption{(Color online). The spin-up or down spectral function
of an asymmetric SIAM
calculated by the HEOM approach at different truncation tiers. The
inset magnifies the Kondo resonance peak at $\w = 0$. The parameters
adopted are $\epsilon_{\mathrm{d}} = -5$, $U = 15$,
$W = 10$, and $T = 0.075$ (in unit of $\Delta$).} \label{fig1}
\end{figure}

It is emphasized that the HEOM approach is distinctly different from
the conventional equations of motion (EOM) method using many-body
Green's functions (GFs) as basic variables \cite{Luo999710}.
The GF--EOM method often treats the impurities and reservoirs on equal
footing. To close the equations it invokes specific approximations for
individual GFs.
In contrast, the HEOM approach focuses on the impurities, with all
reservoir degrees of freedom averaged out properly \cite{Jin08234703}.
Consequently, the HEOM involve much fewer unknowns than GF--EOM at same
tier level.
Moreover, the generic form of \Eq{HEOM} applies to any complex impurity
system, without additional derivation effort. Therefore, the HEOM
approach outperforms GF--EOM in terms of both efficiency and
universality \cite{Sup1}.

For numerical demonstrations, consider first an asymmetric
single-impurity Anderson model (SIAM) system that has been widely
studied \cite{Hew93}. $H_{\s}=\epsilon_{\rm d}(\hat n_{\up}+\hat
n_{\down}) +U\hat n_{\up}\hat n_{\down}$, where $\hat n_{\mu}=\hat
a^{\dg}_{\mu} \hat a_{\mu}$ and $U \neq -2\,\epsilon_{\mathrm d}$.
Lorentzian hybridization functions,
$\Delta_{\mu\nu}(\omega)=\delta_{\mu\nu}\Delta
W^{2}/(\omega^{2}+W^{2})$, are adopted, with $\Delta$ being the
effective impurity-reservoir coupling strength and $W$ the reservoir
band width.
Figure~\ref{fig1} depicts the calculated impurity spectral function
$A(\w)$ by the HEOM approach, up to the converged tier level. The
well-known spectral features of SIAM are clearly resolved:
({\it i}) The two resonance peaks at around $\w =
\epsilon_{\mathrm{d}}$ and $U+\epsilon_{\mathrm{d}}$ correspond to the
excitation energies associated with change of impurity occupancy state.
({\it ii}) The peak at the Fermi energy ($\w = E_F \equiv 0$)
highlights the presence of Kondo resonance under a low temperature.
({\it iii}) The sum rule $\int A(\omega)\,
d\omega=1$ is satisfied to numerical precision.
The comparison in \Fig{fig1} demonstrates distinctly that the HEOM
results converge rapidly with $L$ for full energy range.
This confirms that the HEOM results converge quantitatively at a
relatively low truncation level, even in the Kondo regime.

\begin{figure}
\includegraphics[width=0.85\columnwidth]{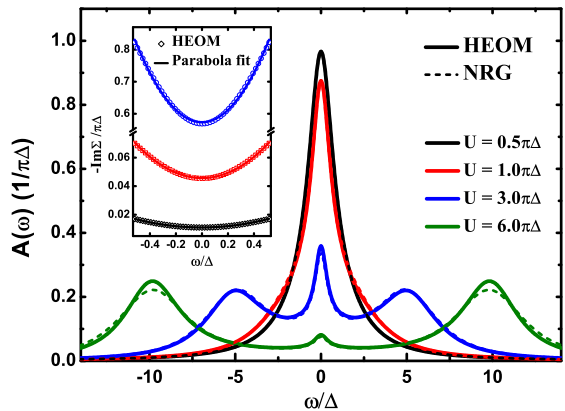}
\caption{(Color online). Comparison between $A(\w)$ of symmetric SIAM
calculated by HEOM and NRG methods. The parameters adopted are $T=0.2$
and $W=50$ (in unit of $\Delta$).
The inset shows the imaginary part of interaction
self-energy calculated from HEOM at energy close to $\omega=0$. }
\label{fig2}
\end{figure}

Figure \ref{fig2} depicts the calculated $A(\w)$ of a symmetric ($U = -
2\epsilon_{\mathrm d}$) SIAM, from weak ($U=0.5\pi\Delta$) to strong
($U=6\pi\Delta$) \emph{e-e} interactions. For comparison, we also show
results obtained by using the full density matrix NRG method
\cite{Wei07076402}, where a self-energy scheme of
Ref.\,[\onlinecite{Bul988365}] is employed, and the results are
averaged over 8 different logarithmic discretizations
\cite{Oli9411986}.
Note that our NRG data in \Fig{fig2} differ slightly from those in
Ref.\,[\onlinecite{Isi10235120}], due to different
$\Delta_{\mu\nu}(\w)$ used (Lorentzian versus constant).
Apparently, the two sets of curves agree quantitatively at all values
of $U$ studied. In the weak ($U=0.5$ and $1.0\,\pi\Delta$) and
intermediate ($U=3\pi\Delta$) interaction regimes, HEOM and NRG curves
almost overlap with each other; while in the strong ($U=6\pi\Delta$)
interaction regime minor deviation is observed in the height of Hubbard
peaks, which is possibly due to remaining uncertainty in NRG results
\cite{Sup1}.
Therefore, such a benchmark comparison clearly affirms that the HEOM
approach achieves the same level of accuracy as the latest high-level
NRG method.

Highlighted in the inset of \Fig{fig2} are the imaginary part of
interaction self-energy (circles), exhibiting a parabolic lineshape
near $\omega=E_F\equiv 0$ (lines). This is a clear indication of Fermi
liquid character \cite{Hew93}. Luttinger has proved that the Kondo peak
height is exactly $1/\pi\Delta$ for a symmetric SIAM at $T=0$,
independent of $U$ \cite{Lut61942}. At finite $T$ and $U$, it is
expected that in general $A(\omega=0)<1/\pi\Delta$ \cite{Yam75970}, as
exemplified by both \Fig{fig1} and \Fig{fig2}.

\begin{figure}
\includegraphics[width=0.95\columnwidth]{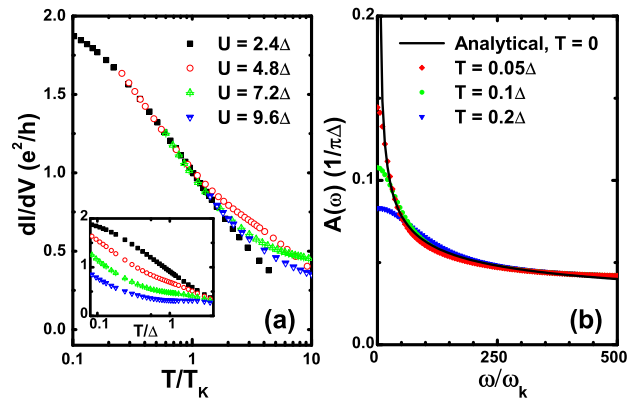}
\caption{(Color online). (a) $dI/dV$ versus $T/T_K$ for symmetric SIAM
with $T_K = (U\Delta/2)^{\frac{1}{2}}\,e^{-\pi U/8\Delta + \pi\Delta/2U}$ \cite{Hew93}
and $W = 24\Delta$.
The inset depicts $dI/dV$ versus unscaled $T$.
(b) Comparison between the HEOM numerical results
and an analytical expression, Eq.~(4.2) of Ref.\,[\onlinecite{Dic014505}],
for the large--$\w$ tail of $A(\w)$.
Other parameters adopted are (in unit of $\Delta$): $W=100$,
$U=-2\ep_{\rm d}=6\pi$. See Ref.\,[\onlinecite{Sup1}] for more details.}
\label{fig3}
\end{figure}

The HEOM results exhibit the correct scaling behavior by Kondo
temperature $T_K$ \cite{Kri801003, Kri801044}. This is verified by the
calculated $dI/dV$ versus $T/T_K$ as depicted in \Fig{fig3}(a), where
the universal scaling is clearly manifested at $T < T_K$ and $\Delta
\ll U \ll W$. Moreover, as $T$ is lowered, the calculated $A(\w)$ draws
progressively to an analytic curve of a logarithmic form predicted in
Ref.\,[\onlinecite{Dic014505}]; see \Fig{fig3}(b).

We also compare HEOM with the latest continuous time QMC (CTQMC)
approach \cite{Gul11349} on the SIAM studied in \Fig{fig2}. Both
approaches yield quantitatively consistent imaginary time GFs with a
maximum relative deviation less than $5\%$. However, $A(\omega)$ of
CTQMC suffer from nontrivial uncertainties in analytical continuation
of GFs to real energies by the maximum entropy method \cite{Sup1}.
We then extend the comparison to the exact diagonalization
\cite{Dag94763,Caf941545,Si942761}, the slave-boson mean-field theory
\cite{Kot861362}, and the non-crossing approximation \cite{Bic87845}.
The HEOM approach is apparently much more accurate than these methods
\cite{Sup1}.
In contrast to the fact that some existing methods would encounter
practical or intrinsic problems in treating certain forms of \emph{e-e}
interactions, the HEOM approach admits an arbitrary form of \emph{e-e}
interaction (including spin-flip, electron-pair hopping, and nonlocal
Coulomb interaction \cite{Hub64237}) without additional computational
cost, as long as it works with the full impurities Fock space.

The computational cost (time and memory) of present HEOM approach grows
rapidly with the lowered $T$. This is because the resolution of
long-time memory requires more exponential functions, and a higher $L$
is usually necessary to achieve quantitative convergence.
In particular, the cost for producing $A(\omega)$ of \Fig{fig2} is
comparable to that required for NRG and CTQMC \cite{Sup1}; while at a
higher $T$, the HEOM approach would be orders of magnitude faster.
Whereas at extremely low $T$ or large $\Delta$, the present HEOM
approach may be very expensive.
It is however possible to reduce the computational cost significantly
by designing more efficient reservoir memory decomposition schemes.

\begin{figure}
\includegraphics[width=0.95\columnwidth]{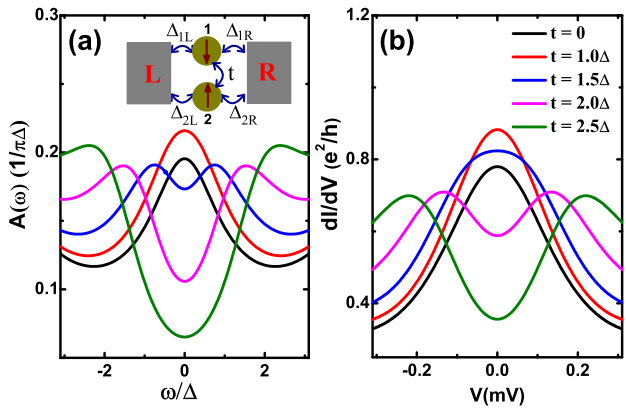}
\caption{(Color online). (a) $A(\w)$
and (b) $dI/dV$ versus $V$ of a TIAM at various inter-impurity coupling
strength ranging from $0$ to $2.5\Delta$. The TIAM system
is sketched in (a). The parameters adopted are
(in unit of $\Delta$): $W=10$, $U_{1}=U_{2}=10$,
$\epsilon_{1}=\epsilon_{2}=-5$, and $T=0.5$.
}
\label{fig4}
\end{figure}

We proceed to demonstrate that the applicability of HEOM approach can
be extended beyond the simple SIAM model and equilibrium properties. To
this end, a parallel-coupled two-impurity Anderson model (TIAM)
sketched in \Fig{fig4}(a) is considered, where $H_{\s} = H_1 + H_2 +
V_{12}$, with $H_1$ ($H_2$) being the SIAM Hamiltonian for the impurity
1 (2), and $V_{12} = t(\hat{a}_{1\uparrow}^\dag \hat{a}_{2\uparrow}+
\hat{a}_{1\downarrow}^\dag \hat{a}_{2\downarrow} + {\rm H.c.})$.
Such a TIAM model has been realized experimentally via a double quantum
dot system, with the inter-dot coupling strength $t$ tuned by plunger
gates \cite{Che04176801}.
The nonzero $t$ gives rise to an effective anti-ferromagnetic coupling,
$J = 4t^2/U$, between the local spin moments at the two impurities. At
a weak $J$, the two spin moments are nearly independent of each other,
and the local spin at each impurity is screened by itinerant electrons
separately. In contrast, at a sufficiently strong $J$, singlet
spin-states covering both impurities are formed.
Therefore, by varying the strength of $t$, the TIAM undergoes a
continuous transition from Kondo singlet states of individual impurity
to singlet spin-states formed between two impurities, as confirmed by
NRG and conformal-field-theory calculations \cite{Sak9081,Aff959528}.

The HEOM approach is applied to evaluate the equilibrium spectral
function $A(\omega)$ of a TIAM consisting of two identical impurities,
along with its $dI/dV$ versus external bias $V$. The latter is a
nonequilibrium property, and is achieved via a finite difference
approach \cite{Sup1}. The response current is extracted from first-tier
($n=1$) auxiliary density operators \cite{Jin08234703}.
Figure~\ref{fig4}(a) and (b) depict the calculated $A(\omega) \equiv
A_1(\omega) = A_2(\omega)$ and $dI/dV-V$, respectively.
Apparently, the variation of $A(\w)$ and $dI/dV-V$ with increasing $t$
are analogous to each other. Their common features are as follows.
(\emph{i}) The system undergoes a transition from a Kondo singlet
involving individual impurity (characterized by the single-peaked
lineshape) at $t < \Delta$, to the singlet spin-state between two
impurities (characterized by the double-peaked lineshape) at $t >
1.5\,\Delta$.
(\emph{ii}) The transition exhibits continuous crossover. As $t$
increases, the single Kondo peak first broadens and approaches to its
maximal height before it drops and splits into two.
These features are consistent with previous experimental
\cite{Che04176801} and theoretical \cite{Sak9081,Aff959528}
investigations.

To summarize, the practicality of our developed hierarchical
Liouville-space approach is demonstrated through studies
on Anderson impurity models, where the key Kondo resonance and Fermi liquid
features due to strong \emph{e-e} interaction are accurately characterized.
The HEOM approach can be straightforwardly extended to more complex
quantum impurity models (such as multi-impurity models) without
additional derivation and programming efforts \cite{Sup1}.
Once converged, the HEOM results can serve as benchmarks to
calibrate approximate numerical approaches, particularly the
effective single-electron approaches, which are useful for studying
more complex systems.
Moreover, it is anticipated that HEOM would become a promising
impurity solver for characterizing strongly correlated lattice
systems in the framework of dynamical mean field theory
\cite{Geo9613}.

%
%
The support from the Hong Kong UGC (AoE/P-04/08-2)(YJY), the NSF of
China (Nos.\,11074302, 11074303, 21103157, 21033008, 21233007), the
Fundamental Research Funds for Central Universities of China
(Nos.\,2340000034, 2340000025)(XZ), the Research Funds of Renmin
University of China (No.\,11XNJ026)(JHW), and the China NKBRFSC
(No.\,2012CB921704)(NHT) is gratefully appreciated. We thank P. Werner,
W. Wu, L. Huang, L. Du, and X. Dai for their help on CTQMC
calculations.
%




\end{document}


\title{Hierarchical Liouville-space approach for
  accurate and universal characterization of quantum impurity systems}

\author{ZhenHua~Li}
\affiliation{Department of Physics, Renmin University of China,
Beijing 100872,
 China }

\author{NingHua~Tong}
\affiliation{Department of Physics, Renmin University of China,
Beijing 100872, China }

\author{Xiao~Zheng}\email{xz58@ustc.edu.cn}
\affiliation{Hefei National Laboratory for Physical Sciences at the
Microscale, University of Science and Technology of China, Hefei, Anhui
230026, China}

\author{Dong~Hou}
\affiliation{Hefei National Laboratory for Physical Sciences at the
Microscale, University of Science and Technology of China, Hefei, Anhui
230026, China}

\author{JianHua~Wei}\email{wjh@ruc.edu.cn}
\affiliation{Department of Physics, Renmin University of China,
Beijing 100872, China }

\author{Jie~Hu}
\affiliation{Department of Chemistry, Hong Kong University of Science
and Technology, Hong Kong, China}
\affiliation{Department of Physics, Capital Normal University, Beijing
100048, China}

\author{YiJing~Yan} \email{yyan@ust.hk}
\affiliation{Hefei National Laboratory for Physical Sciences at the
Microscale, University of Science and Technology of China, Hefei, Anhui
230026, China} \affiliation{Department of Chemistry, Hong Kong
University of Science and Technology, Hong Kong, China}

\date{\today}

\noindent {\bf Supplemental Materials for}

\maketitle

\tableofcontents

\clearpage

\renewcommand{\theequation}{S\arabic{equation}}
\renewcommand{\thetable}{S\arabic{table}}
\renewcommand{\thefigure}{S\arabic{figure}}

\section{General remarks on the HEOM formalism}
\label{thS1}

The hierarchical equations of motion (HEOM) formalism is established
based on the Feynman-Vernon influence functional path-integral
theory,\cite{Fey63118,Kle09,Xu05041103,Xu07031107} and implemented with
the Grassmann algebra for fermionic
dissipatons.\cite{Kle09,Jin08234703}
Thus, it is formally exact
for a general open system coupled to reservoirs bath which satisfies
Grassmann Gaussian statistics.
The mathematical construction of HEOM
has been discussed comprehensively in
Refs.~\onlinecite{Xu05041103,Xu07031107,Jin08234703,Zhe09164708}.
We shall not repeat the tedious derivations, but just address some key
features of the HEOM formalism, particularly for those pertinent to the
quantum impurity systems studied in the main text of this work.
By elucidating these basic features we will discuss the reasons why the
HEOM formalism is not just a formally exact theory, but also leads to
an accurate, efficient and universal approach for characterization of
quantum impurity systems at finite temperatures.

\subsection{Accuracy and efficiency of the HEOM approach}
\label{thS1A}

The efficiency of the HEOM approach is rooted mainly at its
nonperturbative nature, as discussed later. As a result, it converges
often at a relatively low tier level ($L\lesssim 4)$. This appealing
feature is related to the fact that the HEOM formalism is of finite
support, as seen below.

($\bm i$) The HEOM formalism is of finite support by a maximum tier
level $L_{\rm max}$, although numerically it is usually too high to
reach.
Let us start with the HEOM, Eq.~(1) of main text, in which the basic
variables are a set of ADOs, $\{\rho^{(n)}_{j_1\cdots j_n}(t);
n=0,1,\cdots\!,L\}$, with $L$ denoting the converged or truncated tier
level in practice. The reduced system density operator is set to be the
zeroth-order ADO; {\it i.e.}, $\rho^{(0)}(t)\equiv \rho(t) = {\rm
tr}_{\text{res}} [\rho_{\rm total}(t)]$, with ${\rm tr}_\text{res}$
representing the trace over all reservoir environment degrees of
freedom. In constructing HEOM, the influence of reservoirs enters via
the reservoir correlation functions $\{C^{\pm}_{\mu\nu}(t)\}$, which
are related to the inverse Laplace-Fourier transform of the self-energy
functions via $C^{+}_{\mu\nu}(t)=i[\Sigma^{<}_{\mu\nu}(t)]^{\ast}$ and
$C^{-}_{\mu\nu}(t)=i\Sigma^{>}_{\mu\nu}(t)$.
To achieve formally close HEOM, $\{C^{\pm}_{\mu\nu}(t)\}$ are
expanded in exponential series.
Without loss of generality, consider $C^{\sigma}_{\mu\nu}(t)=
\delta_{\mu\nu}\sum_{m=1}^{M}\eta^{\sigma}_{\mu m}
e^{-\gamma^{\sigma}_{\mu m}t}$, with the total number of distinct
exponential terms being $K=2MN_{\mu}$, where the factor $2$ comes from
the choice of $\sigma = +$ or $-$, $M$ is determined by resolution of
reservoir memory, and $N_{\mu}$ denotes the number of system states
through which the itinerary electrons transferring into ($\sigma=+$)
and out of ($\sigma=-$) the central system. In writing the HEOM (1) and
the involving ADOs, we adopt the abbreviation $j\equiv (\sigma\mu m)$,
so that $\gamma_{j}\equiv \gamma^{\sigma}_{\mu m}$. Thus, the damping
term in Eq.\,(1) collects all involving exponents. Denote also $\bar
j\equiv (\bar\sigma\mu m)$, with $\bar\sigma$ being the opposite sign
of $\sigma$.
Involved are also the superoperators,
${\cal A}_{\bar j}\equiv {\cal A}^{\bar\sigma}_{\mu}$
and ${\cal C}_j\equiv {\cal C}^{\sigma}_{\mu m}$,
defined via their actions on an arbitrary operator  of
fermionic or bosonic nature, $\hat O^{\text{\tiny F}}$ or
 $\hat O^{\text{\tiny B}}$, respectively, by
\bsube\label{calAC_Sch}
\be
 {\cal A}^{\sigma}_{\mu} \hat O^{\text{\tiny F}}\equiv
  [\hat a^{\sigma}_{\mu}, \hat O^{\text{\tiny F}}]
\ \ \text{and} \ \
{\cal C}^{\sigma}_{\mu m} \hat O^{\text{\tiny F}}\equiv
  \eta^{\sigma}_{\mu m}\hat a^{\sigma}_{\mu}\hat O^{\text{\tiny F}}
  +(\eta^{\bar\sigma}_{\mu m})^{\ast}\hat O^{\text{\tiny F}}\hat a^{\sigma}_{\mu},
\ee
or
\be
 {\cal A}^{\sigma}_{\mu} \hat O^{\text{\tiny B}}\equiv
   \{\hat a^{\sigma}_{\mu}, \hat O^{\text{\tiny B}}\}
\ \ \text{and} \ \
{\cal C}^{\sigma}_{\mu m} \hat O^{\text{\tiny B}}\equiv
  \eta^{\sigma}_{\mu m}\hat a^{\sigma}_{\mu}\hat O^{\text{\tiny B}}
  -(\eta^{\bar\sigma}_{\mu m})^{\ast}\hat O^{\text{\tiny B}}\hat a^{\sigma}_{\mu}.
\ee
\esube
In particular, while the reduced system density operator
$\rho\equiv \rho^{(0)}$ and its associated even-tier ADOs  $\big\{\rho^{(2k)}\big\}$
are bosonic, the odd-tier $\big\{\rho^{(2k+1)}\big\}$ are fermionic.

Moreover, the subscription index set ($j_1\cdots j_n$) in a generic
$n^{\rm th}$-tier ADO $\rho^{(n)}_{j_1\cdots j_n}$ belongs to an
ordered set of $n$ {\it distinct} $j$-indices. Swapping any two of them
leads to a minus sign to the ADO, \emph{i.e.},
$\rho_{j_rj_p}=-\rho_{j_pj_r}$. 
As a result, the number $K$ of distinct $j$-indices is equal to not
only the number of the first-tier ADOs in total, but also the maximum
hierarchical level ($L_{\rm max}=K$) as by the full HEOM theory. 
Apparently, the number of the $n^{\rm th}$-tier ADOs is
$\frac{K!}{n!(K - n)!}$,
while the total number of unknowns
to solve, up to the truncated tier level $L$,
is $\sum_{n=0}^L \frac{K!}{n\,!(K - n)\,!} \leq 2^K$,
as $L\leq K$.
The overall computational cost increases dramatically with both $L$ and $K$.

To minimize the computational expenditure while maintaining the
quantitative accuracy, various reservoir spectrum decomposition schemes
(in relation to the $K$ size)
have been developed, including the Matsubara spectrum decomposition,\cite{Jin08234703}
the hybrid Matsubara decomposition-frequency dispersion scheme,\cite{Zhe09164708}
the partial fractional decomposition,\cite{Cro09073102} and the
recently proposed Pad\'{e} spectrum
decomposition (PSD) scheme.\cite{Hu10101106,Hu11244106}
Among all these schemes, the PSD scheme provides an optimal basis to
exploit the different characteristic time scales associated with
system-reservoir dissipation processes, and hence has the best
performance. It leads to an optimal construction of HEOM, with the
minimal size of $K$.

($\bm{ii}$) The HEOM formalism is nonperturbative.
 An important feature of the formalism is that for noninteracting
electronic systems (\emph{e.g.} the single-impurity Anderson model
(SIAM) with $U = 0$) that are characterized completely by
single-particle properties, the resultant HEOM terminate automatically
at second tier level ($L=2$) without approximation.\cite{Jin08234703}
This feature highlights the nonperturbative nature of the HEOM
approach.
The underlying hierarchy construction resolves nonperturbatively the
combined effects of \emph{e-e} interaction, system-reservoir
dissipative couplings, and non-Markovian memory of reservoir. For
quantum impurity systems with nonzero \emph{e-e} interactions
(\emph{e.g.} the SIAM with $U \neq 0$), in principle the hierarchy
extends to the $K$-tier level, as $L_{\rm max} =K$ discussed
earlier. In practice, the calculation results usually converge
quantitatively at a low truncation tier ($L \leq 4$), for weak to
medium system-reservoir coupling strength.
It is important to verify the numerical convergence with the
increasing $L$, and such testing procedure has been carried out for
all calculations presented in the main text and also in the
Supplemental Materials.

The above features imply the following analogies of the HEOM formalism
to conventional many-particle quantum mechanics theories. The full HEOM
theory (with $L_{\rm max}=K$) having a total of $2^K$ unknowns
resembles the $K$-particle full-space configuration interaction theory,
but now for $K$ dissipatons resulted from decomposition of reservoir
correlation functions. On the other hand, practically the HEOM
formalism resembles a certain coupled-cluster method, as it converges
rapidly at a relatively low truncation tier level.

\subsection{A universal quantum mechanical method for open systems}
\label{thS1B}

The versatility of the HEOM formalism is rooted at the fact that
reduced system dynamics is completely representable in a linear space,
referred as the HEOM space hereafter. Such a linear space can be deemed
as an extension of conventional Liouville space to open systems.
In this subsection, we will establish the HEOM space and elaborate its
associated algebra. We will demonstrate that the HEOM space is
naturally compatible with the \Sch picture, the Heisenberg picture, and
the interaction picture of the HEOM formalism.
Linear response theory can be established in the HEOM space for open
quantum systems. Its application to the evaluation of system
correlation functions and response functions will be presented in
\Sec{thS1C}. In particular, the calculation of various dynamical
properties of quantum impurity systems, such as the spectral function,
dynamic admittance, and local magnetic susceptibility, will be
discussed in \Sec{thS1D}.

\subsubsection{The \Sch picture versus the Heisenberg picture}
\label{thS1B1}

Let us start with the Schr\"{o}dinger/Heisenberg picture
in the conventional Liouville (or Hilbert) space of isolated systems.
Consider the expectation value of a system dynamical variable,
\be\label{barA0}
  \bar A(t) = {\rm tr}[\hat A\rho(t)]\equiv \la\la \hat A|\rho(t)\ra\ra
 = \la\la \hat A(t)|\rho(0)\ra\ra .
\ee
The last two quantities are the evaluations
in the \Sch picture and the Heisenberg picture, respectively.
The former is described by the Liouville-von Neumann equation,
$\dot\rho=-i{\cal L}\rho = -i[H_{\s},\rho]$, for isolated systems
with $\rho(t)={\cal G}(t,\tau)\rho(\tau)$.
The Heisenberg picture is usually defined only for time-independent
system Hamiltonian $H_{\s}$ cases. The corresponding Liouville-space
propagator of the isolated system is of the translational invariance in
time and given by ${\cal G}(t,\tau)= {\cal G}(t-\tau)=e^{-i{\cal
L}(t-\tau)}$. The Heisenberg picture is then $\hat A(t)=\hat A{\cal
G}(t)$ or equivalently $\frac{\partial}{\partial t}\hat A=-i\hat A{\cal
L}= -i[\hat A, H]$ in isolated systems.

  To identify the HEOM space and the corresponding Heisenberg picture,
  we recast \Eq{barA0} with its HEOM-space evaluation,
\be\label{barA_HEOM}
  \bar A(t) = \la\la \hat{\bm A}(0)|{\bm\rho}(t)\ra\ra
   = \la\la \hat{\bm A}(t)|{\bm\rho}(0)\ra\ra .
\ee
The above identities go by
\be\label{brho}
  {\bm \rho}(t) \equiv \left\{\rho_{j_1\cdots j_n}^{(n)}(t);
 \; n=0,1,\cdots\!,L \right\} ,
\ee
and
\be\label{bA}
  \hat{\bm A}(t) \equiv \left\{\hat A^{(n)}_{j_n\cdots j_1}(t); \;
  n=0,1,\cdots\!,L \right\}\, ,
\ee
with the (by-definition) initial conditions of
$\hat A^{(0)}(t=0)\equiv \hat A$ and $\hat A^{(n>0)}_{j_n\cdots j_1}(t=0)\equiv 0$.
The HEOM-space inner product is defined by
\be\label{inner_product}
 \la\la {\bm A}|{\bm B}\ra\ra \equiv
  \sum_{n=0}^{L} \sum_{j_1\cdots j_n}
   \la\la {A}^{(n)}_{j_n\cdots j_1}|B^{(n)}_{j_1\cdots j_n} \ra\ra \,.
\ee
The $\la\la\,\cdot\,|\,\cdot\,\ra\ra$ on the right-hand-side (rhs) has
been given by the second identity of \Eq{barA0}. The bra $\la\la{\bm
A}|$ has the involving operators in a row vector, while the ket $|{\bm
B}\ra\ra$ a column vector. Consequently, the $j$-indexes set
($j_n\cdots j_1$) in the former is of the reversed order from the
($j_1\cdots j_n$) in the latter.

 Let us write the \Sch picture of the HEOM formalism as
\be\label{HEOM_Sch0}
  \dot{\bm\rho}(t)= - i\bfL \bm{\rho}(t).
\ee
It is just the matrix form of HEOM (1); i.e.,
\begin{align}\label{HEOM_Sch}
   \frac{\partial}{\partial t}\rho^{(n)}_{j_1\cdots j_n} =&
   -i\Big({\cal L} -i\sum_{r=1}^n \gamma_{j_r}\Big)\rho^{(n)}_{j_1\cdots j_n}
     -i \sum_{j}\!     
     {\cal A}_{\bar j}\, \rho^{(n+1)}_{j_1\cdots j_nj}
\nl &
    -i \sum_{r=1}^{n}(-)^{n-r}\, {\cal C}_{j_r}\,
     \rho^{(n-1)}_{j_1\cdots j_{r-1}j_{r+1}\cdots j_n} .
\end{align}
The HEOM-space generator ${\bfL}$ is a matrix
of superoperators defined by \Eq{HEOM_Sch}.
It dictates the HEOM-space propagator $\bfG(t)=e^{-i{\bfL}t}$,
for both
 ${\bm\rho}(t)=\bfG(t){\bm\rho}(0)$ and
$\hat{\bm A}(t)=\hat{\bm A}(0)\bfG(t)$.
The HEOM in the Heisenberg picture assumes therefore
\be\label{HEOM_Hei0}
  \frac{\partial}{\partial t}\hat{\bm A}=-i\hat{\bm A}\bfL\, ,
\ee
with the row vector $\hat{\bm A}$ defined in \Eq{bA}.

To have the explicit Heisenberg HEOM expression,
consider the matrix form of \Eq{HEOM_Sch},
with the part involving $\rho^{(n)}_{j_1\cdots j_n}$ explicitly being highlighted
as (denoting ${\cal L}_{{\bf j}_n}
\equiv {\cal L}-i\sum_{r=1}^n \gamma_{j_r}$)
\[
\frac{\partial}{\partial t}
\begin{bmatrix}
  \times \\ {\rho}^{(n-1)}_{j_1\cdots j_{r-1}j_{r+1}\cdots j_n}
\\ {\rho}^{(n)}_{j_1\cdots j_n}
\\ {\rho}^{(n+1)}_{j_1\cdots j_nj}
\\ \times
\end{bmatrix}
= -i
\begin{bmatrix}
   \times & \times & 0  & 0 & 0
\\ \times & \times & (-)^{n-r}{\cal A}_{\bar j_r} & 0 & 0
\\ 0 & (-)^{n-r}{\cal C}_{j_r}  & {\cal L}_{{\bf j}_n} & {\cal A}_{\bar{j}}   & 0
\\ 0 & 0 &   {\cal C}_j  & \times & \times
\\ 0 & 0 & 0  & \times & \times
\end{bmatrix}
\begin{bmatrix}
  \times \\ {\rho}^{(n-1)}_{j_1\cdots j_{r-1}j_{r+1}\cdots j_n}
\\ {\rho}^{(n)}_{j_1\cdots j_n}
\\ {\rho}^{(n+1)}_{j_1\cdots j_nj}
\\ \times
\end{bmatrix}.
\]
In writing the second row of the above equation for
${\rho}^{(n-1)}_{j_1\cdots j_{r-1}j_{r+1}\cdots j_n}$,
we exploit the identity of ${\rho}^{(n)}_{j_1\cdots j_{r-1}j_{r+1}\cdots j_n\,j_r}
=(-)^{n-r}{\rho}^{(n)}_{j_1\cdots j_n}$.
The Heisenberg counterpart is therefore of
\[
\frac{\partial}{\partial t}
\begin{bmatrix}
  \times \\ \hat A^{(n-1)}_{j_n\cdots j_{r+1}j_{r-1}\cdots j_1}
\\ \hat A^{(n)}_{j_n\cdots j_1}
\\ \hat A^{(n+1)}_{jj_n\cdots j_1}
\\ \times
\end{bmatrix}^{T}
= -i
\begin{bmatrix}
  \times \\ \hat A^{(n-1)}_{j_n\cdots j_{r+1}j_{r-1}\cdots j_1}
\\ \hat A^{(n)}_{j_n\cdots j_1}
\\ \hat A^{(n+1)}_{jj_n\cdots j_1}
\\ \times
\end{bmatrix}^{T}
\begin{bmatrix}
   \times & \times & 0  & 0 & 0
\\ \times & \times & (-)^{n-r}{\cal A}_{\bar j_r} & 0 & 0
\\ 0 & (-)^{n-r}{\cal C}_{j_r}  & {\cal L}_{{\bf j}_n} & {\cal A}_{\bar{j}}   & 0
\\ 0 & 0 &   {\cal C}_j  & \times & \times
\\ 0 & 0 & 0  & \times & \times
\end{bmatrix}.
\]
It reads explicitly as
\begin{align}\label{HEOM_Hei}
 \frac{\partial}{\partial t} \hat A^{(n)}_{j_n\cdots j_1}
&= -i\hat A^{(n)}_{j_n\cdots j_1}\Big({\cal L}-i\sum_{r=1}^{n}\gamma_{j_r}\Big)
   -i \sum_j \hat A^{(n+1)}_{jj_n\cdots j_1}{\cal C}_j
\nl\quad &
  -i \sum_{r=1}^n (-)^{n-r}\hat A^{(n-1)}_{j_n\cdots j_{r+1}j_{r-1}\cdots j_1}
  {\cal A}_{{\bar j}_r}.
\end{align}
This is the Heisenberg counterpart of \Eq{HEOM_Sch}.
The Grassmann-parity associated superoperators ${\cal A}_{j}$ and ${\cal C}_j$
act now from the right or backward side on the specified operators that are of either
fermionic or bosonic in nature, dictated by that of $\hat A^{(0)}\equiv \hat A$
and then alternates in every subsequent tier.
We have the following
equivalence to \Eq{calAC_Sch}:
\bsube\label{calAC_Hei}
\be
 \hat O^{\text{\tiny F}}{\cal A}^{\sigma}_{\mu}
 \equiv [\hat O^{\text{\tiny F}},\hat a^{\sigma}_{\mu}]
\ \ \text{and} \ \
 \hat O^{\text{\tiny F}}{\cal C}^{\sigma}_{\mu m} \equiv
  \eta^{\sigma}_{\mu m}\hat O^{\text{\tiny F}}\hat a^{\sigma}_{\mu}
  +(\eta^{\bar\sigma}_{\mu m})^{\ast}\hat a^{\sigma}_{\mu}\hat O^{\text{\tiny F}},
\ee
or
\be
\hat O^{\text{\tiny B}}{\cal A}^{\sigma}_{\mu}
  \equiv \{\hat O^{\text{\tiny B}},\hat a^{\sigma}_{\mu}\}
\ \ \text{and} \ \
\hat O^{\text{\tiny B}}{\cal C}^{\sigma}_{\mu m} \equiv
  \eta^{\sigma}_{\mu m}\hat O^{\text{\tiny B}}\hat a^{\sigma}_{\mu}
  -\hat O^{\text{\tiny B}}\hat a^{\sigma}_{\mu}(\eta^{\bar\sigma}_{\mu m})^{\ast}.
\ee
\esube

\subsubsection{The interaction picture and the Dyson equation}
\label{thS1B2}

  The HEOM space interaction picture can also be readily
established straightforwardly from its Hilbert-space counterpart.
Let us revisit \Eq{HEOM_Sch}, with ${\cal L}+\delta{\cal L}_{\rm pr}(t)$
and thus the HEOM generator ${\bfL}+\delta{\bfL}_{\rm pr}(t)$.
Here, ${\cal L}$ is the impurities system Liouvillian that may
be time-dependent, for example, in the presence of
external pump fields. On the other hand, $\delta{\cal L}_{\rm pr}(t)$ denotes the
additional contribution from the external
``probe'' field interrogation on the system.
Let ${\bfG}_{\rm pr}(t,\tau)$ be the total HEOM propagator
in the presence of the probe field,
while $\bfG(t,\tau)$ be the probe-free counterpart;
i.e., $\frac{\partial}{\partial t} {\bfG}_{\rm pr}(t,\tau)
=-i({\bfL}+\delta{\bfL}_{\rm pr}){\bfG}_{\rm pr}(t,\tau)$
and $\frac{\partial}{\partial t} {\bfG}(t,\tau)
=-i{\bfL}{\bfG}(t,\tau)$,
respectively.
The standard interaction picture technique
leads then readily to the Dyson's equation with the HEOM dynamics the
form of
\be\label{HEOM_DY}
 {\bfG}_{\rm pr}(t,\tau)=
 \bfG(t,\tau)- i\int_{\tau}^{t}\!d\tau' \, {\bfG}_{\rm pr}(t,\tau')
 \delta{\bfL}_{\rm pr}(\tau')\bfG(\tau',\tau),
\ee
where $\tau$ denotes {\it any time before}
the specified probe field interrogation.
In terms of the interrogated
$\bm{\rho}(t)+\delta\bm{\rho}(t)$,
or $\delta\bm{\rho}(t)=\big[{\bfG}_{\rm pr}(t,\tau)-\bfG(t,\tau)\big]\bm{\rho}(\tau)$,
the above Dyson equation can be recast as
(setting $\tau\rightarrow -\infty$)
\be\label{HEOM_DYrho}
\delta{\bm \rho}(t)=-i\int_{-\infty}^{t}\! d\tau' \,
 {\bfG}_{\rm pr}(t,\tau') \,\delta{\bfL}_{\rm pr}(\tau'){\bm \rho}(\tau')
 =-i\int_{-\infty}^{t}\! d\tau' \,
  {\bfG}_{\rm pr}(t,\tau') [\delta{\cal L}_{\rm pr}(\tau'){\bm \rho}(\tau')]\, .
\ee
In writing the second identity, we emphasize the fact
that $\delta{\cal L}_{\rm pr}(t)$ appears in only but every diagonal
matrix element of the HEOM space generator,
so that $\delta{\bfL}_{\rm pr}(t)=\delta{\cal L}_{\rm pr}(t){\bm{\mathcal{I}}}$,
where ${\bm{\mathcal{I}}}$ is the unit matrix.
Hence,
$\delta{\bfL}_{\rm pr}(t){\bm \rho}(t)
=\big\{\delta{\cal L}_{\rm pr}(t)\rho^{(n)}_{j_1\cdots j_n}(t);
\, n=0,\cdots\!,L\big\}=\delta{\cal L}_{\rm pr}(t){\bm \rho}(t)$,
as inferred from \Eq{brho}.

\subsection{Linear response theory via the HEOM dynamics}
\label{thS1C}

 Consider now the equilibrium HEOM dynamics where
the impurities system is initially at thermal equilibrium,
$\bm\rho(t_0)=\bm\rho^{\rm eq}(T)$ at a given temperature $T$,
and there is no time-dependent pump field.
Therefore, the probe-free propagator is of the translational
invariance in time, $\bfG(t,\tau)=\bfG(t-\tau)$,
and further that $\bm{\rho}(\tau')=\bfG(\tau'-t_0)\bm\rho(t_0)=\bm\rho^{\rm eq}(T)$
in the integrand of \Eq{HEOM_DYrho}.
Together with the Dyson equation, \Eq{HEOM_DYrho} for the equilibrium
dynamics reads
\be\label{HEOM_perturbation}
\delta{\bm \rho}(t)
 =\sum_{n=1}^{\infty}
    (-i)^n \int_{-\infty}^{t}\!\! d\tau_n \cdots\!\int_{-\infty}^{\tau_2}\!\!d\tau_1\,
  {\bfG}(t-\tau_n)\delta{\cal L}_{\rm pr}(\tau_n){\bfG}(\tau_n-\tau_{n-1})
 \cdots
   \delta{\cal L}_{\rm pr}(\tau_1)\bm\rho^{\rm eq}(T)\, .
\ee
This resembles the celebrated Hilbert-space time-dependent perturbation
theory via interaction picture.
Note that the thermal equilibrium $\bm\rho^{\rm eq}(T)\equiv
\big\{\rho^{(n);\,{\rm eq}}_{j_1\cdots j_n}(T);
\, n=0,\cdots\!,L\big\}$ is a steady-state solution of the HEOM (\ref{HEOM_Sch}),
in the absence of time-dependent external fields;
see its evaluation in \Sec{numerical}.
The resulting nonzero ADOs, $\rho^{(n>0);\,{\rm eq}}_{j_1\cdots j_n}(T)\neq 0$,
reflect the initial system-reservoirs correlations.

 The above observations highlight a straightforward
equivalence mapping between the conventional Hilbert space
and the HEOM space formulations
for the evaluations of various
correlation and response functions of impurity systems.
Consider, for example, the steady-state two-time correlation function
between two arbitrary dynamical variables $\hat A$ and $\hat B$ of the
impurity system of interest. We have [\emph{cf}.~\Eqs{barA0} and
(\ref{barA_HEOM})]
\be\label{CAB}
 \widetilde{C}_{AB}(t) = \big\la \hat A(t)\hat B(0)\big\ra =
  \big\la\big\la \hat A(t) \big|
   \hat B\rho^{\text{eq}}_{\text{total}}(T)\big\ra\big\ra
 = \big\la\big\la \hat{\bm A}(t) \big|
   \hat B{\bm\rho}^{\text{eq}}(T)\big\ra\big\ra \, .
\ee
Similarly, the response function will be
\be\label{RAB}
  \chi_{AB}(t) = i\, \big\la [\hat A(t),\hat B(0)]\big\ra =
   i \big\la\big\la\hat A(t) \big|
   [\hat B,\rho^{\text{eq}}_{\text{total}}(T)]\big\ra\big\ra
 = i\,\big\la\big\la \hat{\bm A}(t) \big|
    [\hat B,{\bm\rho}^{\text{eq}}(T)] \big\ra\big\ra \, .
\ee
The proof will be exemplified with the response function of \Eq{RAB} as follows.

 Application of \Eq{HEOM_perturbation} is to be illustrated with
the HEOM evaluation of correlation and response functions
of impurity systems via the linear response theory
(LRT) which applies to the HEOM linear space defined in \Sec{thS1B}.

Regarding the LRT, we are primarily interested in how
system properties (such as the expectation value
of a system observable $\hat{A}$) respond to an external
perturbation in the physical subspace of the system.
Suppose the perturbation Hamiltonian assumes the
following form
\begin{equation}
   H_{\rm pr}(t)= - \hat{B} \epsilon_{\rm pr}(t), \label{LRT_HPR}
\end{equation}
where $\hat{B}$ is a Hermitian system operator,
and $\epsilon_{\rm pr}(t)$ is the time-dependent perturbative
field which takes effect from $t_0$ [$\ep_{\rm p}(t<t_0) = 0$].
The induced change in the expectation value of a
system observable $\hat{A}$ due to the presence of $H_{\rm pr}(t)$ is
\be \delta \hat{A}(t) = {\rm tr}_{\s} \big[ \hat{A}\, \delta\rho(t)
\big] = \big\langle \big\langle \hat{\bm A}| \delta{\bm \rho}(t)
\big\rangle \big\rangle. \label{LRT_DA} \ee
By using the first order perturbation in \Eq{HEOM_perturbation}, we
arrive at
\begin{equation}
\delta\boldsymbol{\rho}(t)= -i\int_{0}^{t}d\tau\,
\boldsymbol{\mathcal{G}}(t-\tau)
\delta\boldsymbol{\mathcal{L}}_{\textrm{pr}}(\tau)
\boldsymbol{\rho}^{\textrm{eq}}(T).\label{LRT_1STDR}
\end{equation}
To proceed, we define the time-independent HEOM-space superoperator
$\hat{\boldsymbol{\mathcal{B}}}$ as
\begin{equation}
\hat{\boldsymbol{\mathcal{B}}} \equiv
-\delta\boldsymbol{\mathcal{L}}_{\textrm{pr}}(t)
/\epsilon_{\textrm{pr}}(t),\label{LRT_DEFB}
\end{equation}
whose action can be determined as
\begin{equation}
\hat{\boldsymbol{\mathcal{B}}}\boldsymbol{\rho}=[\hat{B},\;\boldsymbol{\rho}].\label{LRT_ACTB}
\end{equation}
Inserting Eq.\eqref{LRT_ACTB} into \eqref{LRT_1STDR}, we finally get
\begin{equation}
\delta\hat{A}(t)=i\int_{0}^{t}d\tau\, \big\langle \big\langle
\hat{\boldsymbol{A}}(0)|\boldsymbol{\mathcal{G}}(t-\tau)
\hat{\boldsymbol{\mathcal{B}}}|\boldsymbol{\rho}^{\textrm{eq}}(T)
\big\rangle \big\rangle \,
\epsilon_{\textrm{pr}}(\tau),\label{LRT_A2}
\end{equation}
from which we are ready to define the response function in HEOM space as
\begin{equation}
\chi_{AB}(t,\tau)\equiv i\, \big\langle \big\langle
\hat{\boldsymbol{A}}(0)|\boldsymbol{\mathcal{G}}(t-\tau)
\hat{\boldsymbol{\mathcal{B}}}|
\boldsymbol{\rho}^{\textrm{eq}}(T)\big\rangle
\big\rangle.\label{LRT_RAB}
\end{equation}
The correlation function in HEOM space can be similarly defined as
\begin{equation}
\widetilde{C}_{AB}(t,\tau)\equiv \big\langle \big\langle
\hat{\boldsymbol{A}}(0)|
\boldsymbol{\mathcal{G}}(t-\tau)\hat{\boldsymbol{B}}|
\boldsymbol{\rho}^{\textrm{eq}}(T)\big\rangle \big\rangle
,\label{LRT_CAB}
\end{equation}
where
$\hat{\boldsymbol{B}}\boldsymbol{\rho}^{\rm eq}(T) = \{
 \hat{B}\rho^{(n);{\rm eq}}; n=0,1,\ldots, L\}$. By setting
$\tau=0$ while noting
$\hat{\boldsymbol{A}}(t)=\hat{\boldsymbol{A}}(0)\boldsymbol{\mathcal{G}}(t)$,
\Eqs{LRT_RAB} and (\ref{LRT_CAB}) recover \Eqs{RAB} and (\ref{CAB})
immediately.

%

The dynamical observables of our primary interest in the present
work are the system correlation function and spectral function,
which associate with each other through the fluctuation-dissipation
theorem (FDT) at thermal equilibrium. Consider, for example,
the retarded and advanced
single-particle Green's functions in terms of the correlation functions,
\be G_{AB}^{r}(t) = -i\theta(t) \big\langle \big\{\hat{A}(t),\:
\hat{B}\big\} \big\rangle =-i\theta(t)
\left[\widetilde{C}_{AB}(t)+\widetilde{C}_{BA}(-t)\right],\label{RGF_def}
\ee
and
\begin{equation}
G_{AB}^{a}(t)=i\theta(-t) \, \big\langle \big\{\hat{A}(t),\: \hat{B}
\big\} \big\rangle = i\theta(-t) \,
\left[\widetilde{C}_{AB}(t)+\widetilde{C}_{BA}(-t)\right].\label{AGF_def}
\end{equation}
Here, the Green's functions in our theory is defined for two arbitrary
operators $\hat{A}$ and $\hat{B}$.
We now focus on cases where $\hat{A}=\hat{B}^\dag$ (\emph{e.g.},
$\hat{A}=a$ and $\hat{B}=a^\dag$ for conventional fermionic Green's
functions). In such circumstances,
$\widetilde{C}^{\ast}_{AB}(t)=\widetilde{C}_{AB}(-t)$ follows by definition.
Using this relation in the Fourier transform of
\Eqs{RGF_def}$-$(\ref{AGF_def}) leads to
$G_{AB}^{a}(\omega)=[G_{AB}^{r}(\omega)]^{\ast}$.
We can further define the spectral function $J_{AB}(\omega)$ as
\be J_{AB}(\omega)\equiv -\frac{1}{\pi}\, \mathbf{\mathrm{Im}}
\left[G_{AB}^{r}(\omega)\right]=-\frac{1}{2\pi
i}\left[G_{AB}^{r}(\omega)-G_{AB}^{a}(\omega)\right].\label{SecF_def}
\ee
Let
$C_{AB}(\omega)\equiv\frac{1}{2}\int_{-\infty}^{\infty}dt\,
e^{i\omega t}\widetilde{C}_{AB}(t)$, which satisfies
 $C_{BA}(-\omega)=e^{-\beta\omega}C_{AB}(\omega)$,
the detailed-balance relation at thermal equilibrium. By combining
\Eqs{RGF_def}$-$(\ref{SecF_def}), we arrive at the FDT of
\begin{equation}
J_{AB}(\omega)=\frac{1}{\pi}\left(1+e^{-\beta\omega}\right)C_{AB}(\omega).
\label{SF_FDT}
\end{equation}
%


%
\subsection{Evaluation of dynamical observables of quantum impurity
systems} \label{thS1D}

In principle, the dynamical correlation of any two system operators
($\hat{A}$ and $\hat{B}$) can be calculated by using the HEOM-space LRT
via \Eq{LRT_CAB}. This allows us to evaluate at quantitative accuracy a
variety of dynamical observables of experimental significance, such as
the Green's functions and spectral function of the system, local
electric and magnetic susceptibilities, and dynamic admittance (inverse
of complex impedance), \emph{etc.}
The calculation of system Green's functions and spectral function will
be presented in \Sec{numerical}.
Here, we take the local magnetic susceptibility as an example to
illustrate how the response properties are evaluated by using the HEOM
approach.

The local magnetic susceptibility $\chi^m$ describes the magnetization
of an impurity in linear-order response to an external magnetic field
applied locally at the impurity site. For simplicity, we consider the
case that the magnetic field is applied only along the $z$-direction,
so is the induced spin polarization for electrons on the impurity.
\begin{equation}\label{Magsus}
M_z(\w)=\chi^m(\w)\, H_z(\w).
\end{equation}
Here, $M_z(\w)$, $H_z(\w)$, and $\chi^m(\w)$ are the local
magnetization, local magnetic field, and local magnetic susceptibility
of frequency $\w$ along $z$-direction, respectively.
Note that $M_z(\w)= \mathcal{F}[\langle \hat{m}_z(t)\rangle]$, with the
local magnetization operator being
\begin{equation}\label{Magnet}
\hat{m}_z = g\mu_B \hat{S}_z = \frac{1}{2} g\mu_B \sum_i \left( a^\dag_{i\uparrow}
a_{i\uparrow} -  a^\dag_{i\downarrow}a_{i\downarrow} \right).
\end{equation}
Here, $g$ is the gyromagnetic ratio, $\mu_B$ the Bohr magnetization,
and $\hat{S}_z$ is the electron spin polarization operator along
$z$-direction for the impurity.

Upon application of an \emph{ac} local magnetic field, $H_z(t)$, the
system perturbation Hamiltonian is
\begin{equation}\label{Magint}
\hat{H}_{\rm pr}(t)=- \hat{m}_z H_z(t).
\end{equation}
Apparently, \Eq{Magint} assumes the same form as \Eq{LRT_HPR}, with
$\hat{m}_z$ playing the role of $\hat{B}$. The physical observable of
primary interest is $\hat{A} = \hat{m}_z$. Based on the HEOM-space LRT,
the local magnetic susceptibility is expressed as
\begin{equation}\label{RMAB}
\tilde{\chi}^m(t) = i\langle [\hat{m}_z(t), \hat{m}_z]\rangle =
i\,\big\la\big\la \hat{\boldsymbol{m}}_{z}(0) \big|
\boldsymbol{\mathcal{G}}(t)
\big| [\hat{m}_{z},{\bm\rho}^{\text{eq}}] \,\big\ra\big\ra \, .
\end{equation}
Equation~\eqref{RMAB} is formally analogous to \Eq{LRT_RAB}, and hence
$\tilde{\chi}^m(t)$ can be readily calculated by solving the quantum
Liouville propagator $\boldsymbol{\mathcal{G}}(t)$ in HEOM space.
The \emph{dynamic} (frequency-dependent) local magnetic susceptibility
is then obtained via Fourier transform, \emph{i.e.}, $\chi^m(\w) =
\mathcal{F}[\tilde{\chi}^m(t)]$.
In particular, the \emph{static} local magnetic susceptibility is just
its zero-frequency component, $\chi^m(\w=0)$.
The above formulas can be easily extended to scenario of noncollinear
spin.

The above procedures are applicable to a variety of dynamical
observables, such as the local electric susceptibility. However, the
evaluation of dynamic admittance is somewhat different. It involves the
calculation of response function, $\tilde{\chi}^G_{\alpha\alpha'}(t) =
-i\langle [\hat{I}_\alpha(t), \hat{N}_{\alpha'}] \rangle$, where
$\hat{I}_\alpha$ and $\hat{N}_{\alpha'}$ are the current operator for
$\alpha$--reservoir and electron number operator for
$\alpha'$--reservoir, respectively. Although $\hat{I}_\alpha$ and
$\hat{N}_{\alpha'}$ are not ``system'' operators,
$\tilde{\chi}^G_{\alpha\alpha'}(t)$ can still be evaluated by making
use of the quantum Liouville propagator in HEOM space. The detailed
formulation and numerical demonstrations will be published elsewhere.

It is important to emphasize that the HEOM approach is capable of
addressing the nonequilibrium electronic dynamics under arbitrary
time-dependent external fields. This thus provides an alternative way
for the evaluation of dynamical observables, \emph{i.e.}, through the
real-time evolution of system reduced density matrix $\rho(t)$ and the
ADOs. The expectation value of physical observable is then extracted
from $\{\rho^{(n)}(t)\}$.
For instance, the dynamic local magnetic susceptibility can be obtained
via $\chi^m(\w) = g\mu_B\, \mathcal{F}[S_z(t)] / \mathcal{F}[H_z(t)]$,
where $S_z(t) = \langle \hat{S}_z \rangle = {\rm tr}[\hat{S}_z
\rho(t)]$ is the spin polarization for electrons on the impurity under
the applied magnetic field $H_z(t)$.
Another example is the dynamic admittance, which can be evaluated via
$\chi^G_{\alpha\alpha'}(\w) \equiv
\mathcal{F}[\tilde{\chi}^G_{\alpha\alpha'}(t)] =
\mathcal{F}[I_\alpha(t)]/\mathcal{F}[-eV_{\alpha'}(t)]$, with
$I_\alpha(t)$ and $V_{\alpha'}(t)$ being the time-dependent current and
voltage for $\alpha$-- and $\alpha'$--reservoir, respectively.
Note that $I_\alpha(t)$ is extracted from first-tier ADOs; see
Ref.~[\onlinecite{Zhe08184112}].
In our previous work, we have carried out comprehensive studies on the
dynamic admittance of open impurity systems; see
Refs.~[\onlinecite{Zhe08184112, Zhe08093016}].
Moreover, such an approach allows us to go beyond the linear regime and
explore system properties in far-from-equilibrium situations.
Furthermore, the static response ($\w = 0$) to external perturbation
can be obtained straightforwardly by solving stationary states
corresponding to fixed external fields.
For instance, the differential conductance data ($dI/dV$) displayed in
Fig.\,3 of main text are extracted from steady-state current under
different bias voltages via a finite-field treatment.

To summarize, there are two ways to evaluate the dynamical observables
of quantum impurity systems within the framework of HEOM theory. One is
to calculate the relevant correlation/response function via the
HEOM-space LRT; while the other is to calculate the nonequilibrium
static/dynamic electronic response to applied external fields.
These two ways are completely equivalent in the linear response regime.

\section{Numerical aspects of the HEOM formalism}
\label{numerical}

The numerical implementation of the HEOM method for calculations
of dynamical properties of open electronic system, such as
transient transport current through quantum dot systems under
time-dependent
applied voltages, has been addressed in our previous publications; see
Refs.\,[\onlinecite{Zhe08184112, Zhe08093016, Zhe09124508, Zhe09164708}].
Our recent improvement on numerical algorithms and codes
concerns mainly the follows: (i)
improve the numerical efficiency by making use of various advanced
numerical techniques, so that the quantum impurity systems
in strong correlation regime can be treated accurately with the
presently available computational resources;
and (ii) extend the HEOM method to evaluation of
dynamical observables at equilibrium state.

To reduce the computational cost, we have employed
the Pad\'{e} spectrum decomposition scheme\cite{Hu10101106, Hu11244106}
to construct the HEOM, the quasi-minimal residual (QMR)
method\cite{Fre91315, Fre93470}
to solve the large linear problem for stationary states, and
the fourth-order Runge-Kutta\cite{Pre92} or Chebyshev\cite{Kos882087}
method for the time evolution
of quantum Liouville propagator in HEOM space.
We have also used sparse matrix algorithms
and parallel computing techniques to accelerate the calculations.
The detailed descriptions of our numerical code will be
published elsewhere.
Here, we demonstrate the numerical capability of our
HEOM code through the calculations on the same single-impurity Anderson
model (SIAM) as studied in Fig.\,1 of the main text.
Tables~\ref{table:SIAM_highT} and \ref{table:SIAM_lowT} show the
computational cost and numerical convergence for the SIAM
at relatively high ($T = 1.5\,\Delta$) and low ($T = 0.075\,\Delta$)
temperatures, respectively.
In particular, the low temperature brings the SIAM system into the
strong correlation (Kondo) regime, and the fourth-tier truncation ($L =
4$) predicts the number of occupied electrons with a minor uncertainty
of less than $2 \times 10^{-5}$ electron.

A variety of system dynamical observables, such as the spectral density
function associated with $\mu$th single-electron level of system,
$A_\mu(\w) \equiv J_{\hat{a}_\mu \hat{a}_\mu^\dag}(\w)$, can be
evaluated in two ways: either with a time-domain scheme or by
calculations in the frequency domain.

The time-domain scheme starts with the evaluation of system correlation
functions
$\wti{C}_{\hat{a}_\mu^\dag \hat{a}_\mu}(t)$ and
$\wti{C}_{\hat{a}_\mu \hat{a}_\mu^\dag}(t)$
through the time evolution of the quantum Liouville propagator
$\hat{\boldsymbol{\mathcal{G}}}_{\rm eq}(t)$ at $t > 0$. The spectral
function is then obtained straightforwardly
by a half Fourier transform as follows,
\be
  A_\mu(\w) = \frac{1}{\pi}\, {\rm Re} \left\{
   \int_0^{\infty} dt\,
   \Big\{\wti{C}_{\hat{a}_\mu \hat{a}_\mu^\dag}(t)
   +[\wti{C}_{\hat{a}_\mu^\dag \hat{a}_\mu}(t)]^\ast
   \Big \}\, e^{i\w t}  \right\}. \label{Aw_time}
\ee
The averaged system spectral function is $A(\w) = \frac{1}{N}
\sum_{i=1}^N A_i(\w)$. However, in practice the time-domain scheme can
be very time-consuming, especially for strongly correlated systems at
low temperatures, where $A(\w)$ exhibits prominent Kondo signatures at
around zero frequency.
In such cases, both $\wti{C}_{\hat{a}_\mu^\dag \hat{a}_\mu}(t)$
and $\wti{C}_{\hat{a}_\mu \hat{a}_\mu^\dag}(t)$ decay rather slowly at large
$t$, due to the significant long-time reservoir memory component.
Consequently, to calculate $A(\w)$ at low frequency accurately, it is
necessary to have the correlation functions propagate to a long enough
time.

An alternative scheme is based on the half-Fourier transformed
correlation function via
\begin{align}
  \bar{C}_{AB}(\w) &\equiv \int_0^{\infty} dt\, C_{AB}(t)\, e^{i\w t}
  = \int_0^{\infty} dt\, \langle\langle \bm A |
  \, \hat{\boldsymbol{\mathcal{G}}}_{\rm eq}(t) |
  \bm B_{\rm eq} \rangle \rangle \, e^{i\w t} \nl
  &= \int_0^{\infty} dt \, \langle\langle \bm A |
  e^{(i\w - \hat{\bm \Lambda}) t} | \bm B_{\rm eq} \rangle\rangle \nl
  &= - \langle \langle \bm A | \big(i\w - \hat{\bm \Lambda}\big)^{-1}
  | \bm B_{\rm eq}  \rangle \rangle
  = \langle \langle \bm A | \bm X \rangle \rangle. \label{CABw}
\end{align}
At a fixed frequency $\w$, the HEOM-space vector $\bm X$
is determined by solving the following linear problem
with the QMR algorithm:
\be
   \big(i\w - \hat{\bm \Lambda} \big) \bm X = - \bm B_{\rm eq}.
   \label{linX}
\ee
The system spectral function is then
evaluated via
\be
  A_\mu(\w) = \frac{1}{\pi} \, {\rm Re} \Big[
  \bar{C}_{\hat{a}_\mu \hat{a}_\mu^\dag}(\w) +
  \bar{C}_{\hat{a}_\mu^\dag \hat{a}_\mu}(-\w) \Big].
  \label{Aw_freq}
\ee
Within the frequency-domain scheme, $A(\w)$ is calculated
at one frequency point by solving
\Eqs{linX} and (\ref{Aw_freq}) once.
Therefore, to obtain the whole spectrum of $A(\w)$, a
full scan for the whole frequency domain is needed.

\begin{figure}
\includegraphics[width=0.8\columnwidth]{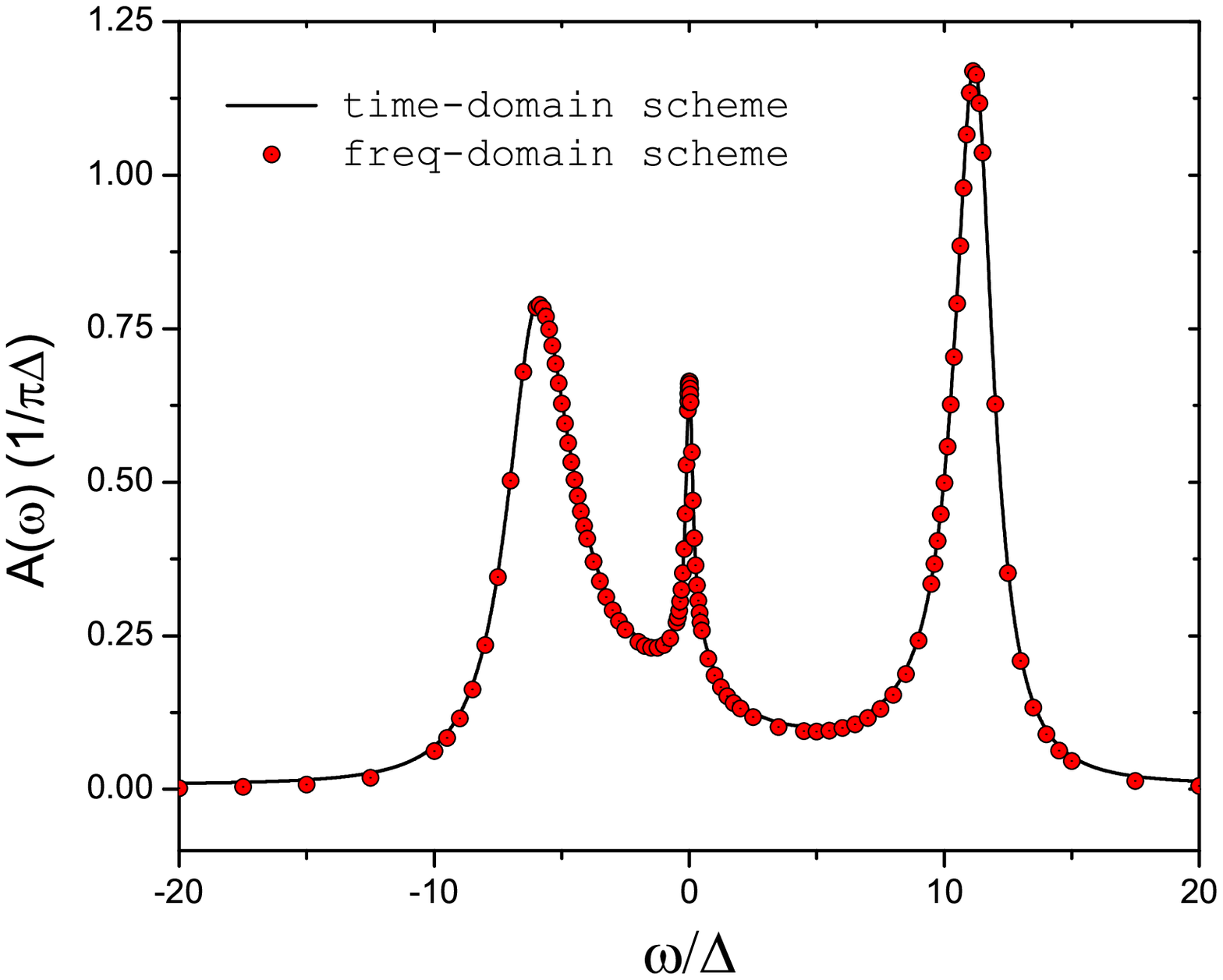}
\caption{Spectral function of an asymmetric SIAM
calculated by the HEOM approach at $L = 4$.
The same set of parameters as that used
for plotting Fig.\,1 in the main text are
adopted here (in unit of $\Delta$):
$\ep_d = -5$, $U = 15$, $W = 10$, and $T = 0.075$.
The results of both time-domain and frequency-domain
schemes are displayed. } \label{figS1}
\end{figure}

The equivalence between the two schemes is exemplified
with \Fig{figS1}, where the spectral function of a
SIAM is calculated with both time- and frequency-domain
schemes. Apparently, the results out of
both schemes agree perfectly with each other, as they should.
In practice, a hybrid time- and frequency-domain scheme can be
employed. The overall profile of $A(\w)$ can be obtained
with the time-domain scheme, while the detailed structures
of the resonance and Kondo peaks which are difficult to
resolve by the time-domain algorithm
can be exploited with the frequency-domain scheme. This hybrid
scheme provides an accurate and efficient method to
calculate the full spectrum of $A(\w)$ for a strongly correlated
quantum impurity system in the Kondo regime.

Similarly, the retarded Green's function is evaluated based
on system correlation functions with the time- or frequency-domain
scheme as follows.
\be
  G^r_{\mu\mu}(\w) =  -i \int_0^{\infty} dt \,
  \Big\{ \wti{C}_{\hat{a}_\mu \hat{a}_\mu^\dag}(t) +
  \big[\wti{C}_{\hat{a}_\mu^\dag \hat{a}_\mu}(t)\big]^\ast
  \Big\}  \,e^{i\w t}
  = -i \left\{ \bar{C}_{\hat{a}_\mu \hat{a}_\mu^\dag}(\w)
  + [\bar{C}_{\hat{a}_\mu^\dag \hat{a}_\mu}(-\w)]^\ast \right\}.
  \label{Grw}
\ee
For an SIAM system, the single-electron retarded
Green's function is formally expressed as
\be
   G^r_d(\w) = [\w - \ep_d - \Sigma^r_{\rm res}(\w)
   - \Sigma^r_{\rm ee}(\w)]^{-1}.  \label{Grdw}
\ee
Here, $\Sigma^r_{\rm res}(\w)$ is the self-energy due to the
dissipative couplings between the impurity and reservoirs.
Its imaginary part gives the hybridization function $\Delta(\w) = -{\rm
Im}[\Sigma^r_{\rm res}(\w)]$, which is calculated from the surface
Green's function of the isolated electron reservoirs, or provided as a
known property of the reservoirs (such as the Lorentzian lineshape
model adopted in this work).
$\Sigma^r_{\rm ee}(\w)$ is the self-energy due to
the presence of electron-electron
interaction (the $U$-term) which reflects the
information of electron correlation.
With $G^r_d(\w)$ evaluated by the HEOM approach
via \Eq{Grw}, $\Sigma^r_{\rm ee}(\w)$ can be
extracted from \Eq{Grdw} as the only unknown variable.

\section{Remarks on application of HEOM approach to
strongly correlated quantum impurity systems} \label{HEOM-app}

\subsection{Application to single-impurity Anderson model} \label{SIAM}

\subsubsection{Spectral function of SIAM}
\label{sub2sec:spectral}

\begin{figure}
\includegraphics[width=0.8\columnwidth]{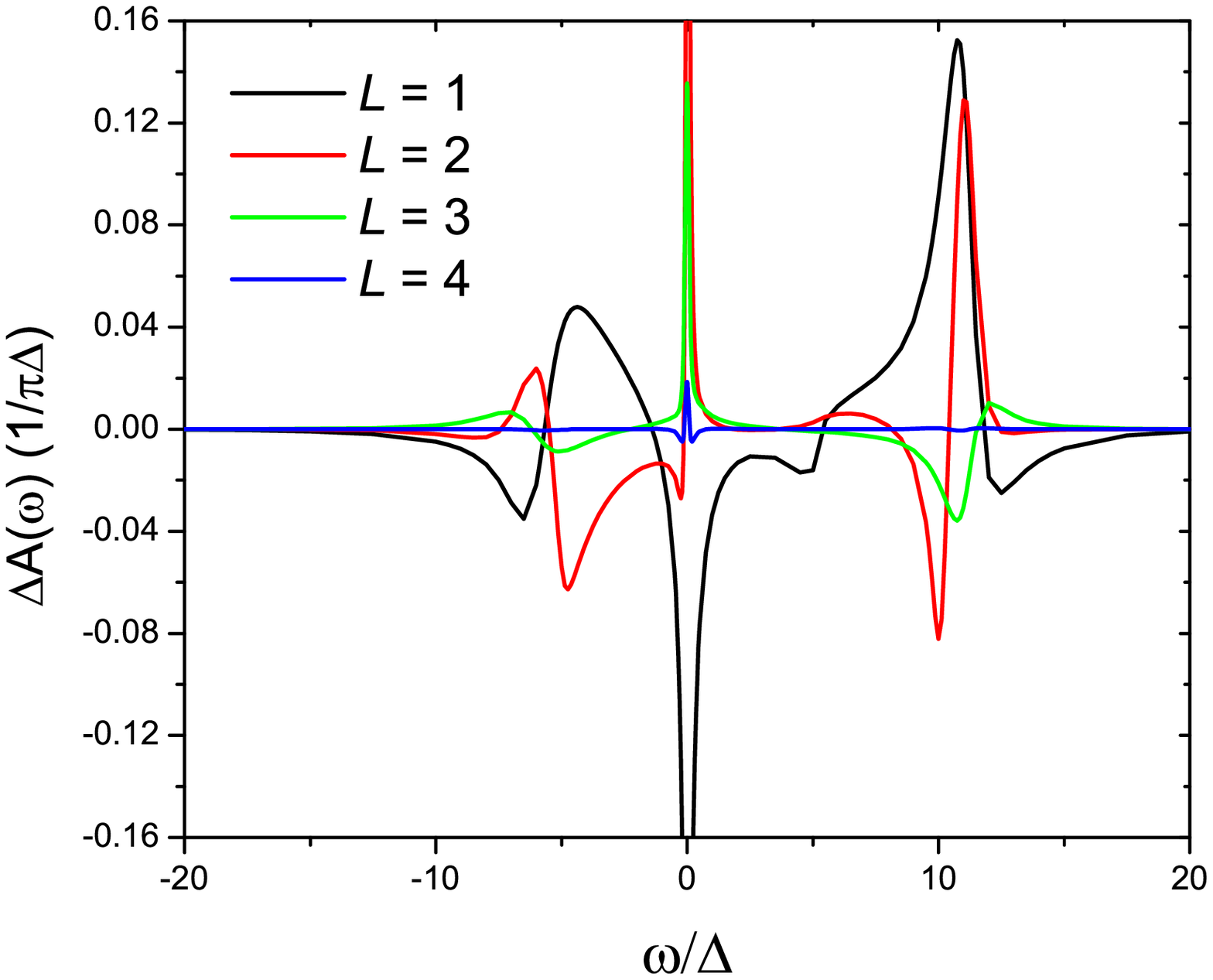}
\caption{Deviation of system spectral function of
an asymmetric SIAM
calculated at $L=1 \sim 4$ from the $L=5$ result.
The same set of parameters as that used
for plotting Fig.\,1 in the main text are
adopted here (in unit of $\Delta$):
$\ep_d = -5$, $U = 15$, $W = 10$, and $T = 0.075$.
} \label{figS2}
\end{figure}

The numerical convergence of HEOM is exemplified via the calculated
$A(\w)$ of an asymmetric SIAM system at different truncation tier $L$;
see Fig.~1 in the main text, where the resulting $A(\w)$ converges
uniformly to the numerically exact solution with the increasing $L$.
Figure~\ref{figS2} depicts the
derivation of $A(\w)$ calculated at various $L$ from the
result obtained at $L_{\rm max} = 5$. It is observed that the
magnitude of difference indeed decreases as $L$ approaches $L_{\rm max}$.
In particular, there is only rather minor deviation between the
calculated $A(\w)$ at $L = 4$ and that at $L = 5$, which is
barely recognizable at around zero frequency.

\begin{figure}
\includegraphics[width=0.8\columnwidth]{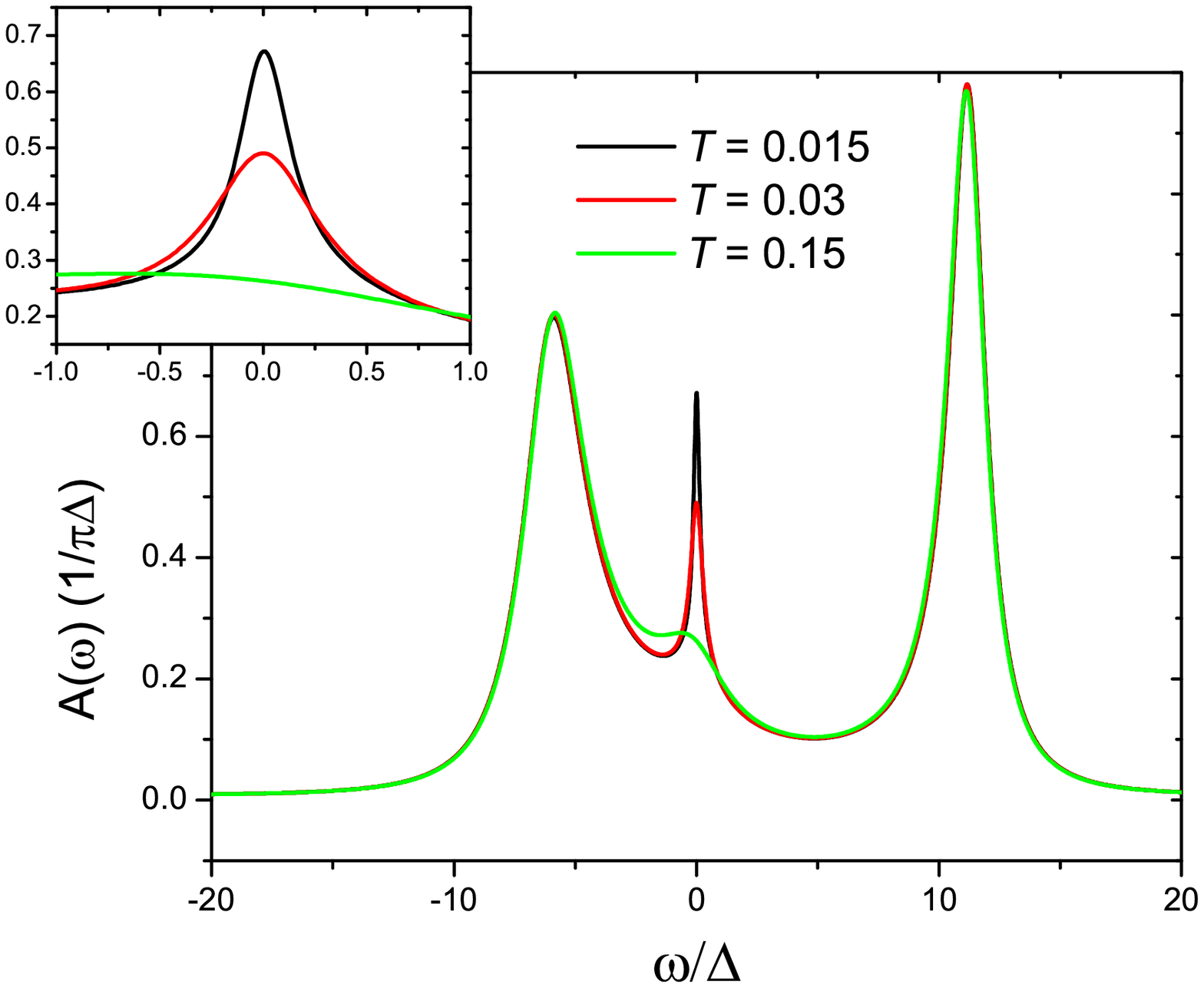}
\caption{Calculated system spectral function of
an asymmetric SIAM at different temperatures.
The same set of parameters as that used
for plotting Fig.\,1 in the main text are
adopted here (in unit of $\Delta$):
$\ep_d = -5$, $U = 15$, $W = 10$, and $T = 0.075$.
All curves are obtained at $L = 5$, and are considered
to be quantitatively converged.
The inset magnifies the low frequency range
where the Kondo resonance peak becomes more
prominent at a lower temperature.
} \label{figS3}
\end{figure}

We also investigate the temperature dependence of $A(\w)$.
Figure~\ref{figS3} plots the calculated $A(\w)$ of the same asymmetric
SIAM system at various temperatures. Apparently, as the temperature
increases over an order of magnitude, the two resonance peaks at $\w =
\ep_d$ and $\w = \ep_d + U$, which are due to transition between
different occupancy states, remain intact. In contrast, the Kondo peak
at $\w = E_{\rm F} = 0$ almost vanishes at the higher temperature. This
clearly highlights the significant role of strong electron correlation
in quantum impurity systems.

\subsubsection{Differential conductance and Kondo temperature}
\label{sub2sec:kondo}

We proceed to examine the zero-bias differential conductance of various
two-terminal SIAM systems. This is to verify that the low-temperature
physical properties obtained by the HEOM approach indeed follows the
correct Kondo scaling relation.

The analytical expression for Kondo temperature of a symmetric SIAM
is\cite{Hew93}
\be
  T_K = \sqrt{\frac{U\Delta}{2}}\, e^{-\pi U/8\Delta + \pi
  \Delta/2U}. \label{eq:TK_symm}
\ee
Equation~\eqref{eq:TK_symm} is derived based on Bethe
Ansatz,\cite{Hew93} and is expected to be valid provided that the
relation $\Delta \ll U \ll W$ is satisfied. Here, $\Delta$ is the
impurity-reservoir coupling strength, and $W$ is the bandwidth of
electron reservoirs.

Figure~\ref{figS4} and Figure~3(a) of main text depict the temperature
dependence of zero-bias differential conductance ($dI/dV$) of symmetric
SIAM systems with different values of $U$. Apparently, the $dI/dV$
versus $T$ curves for different values of $U$ display rather scattered
distribution; see \Fig{figS4}(a). In contrast, with the temperature
scaled by $T_K$ of \Eq{eq:TK_symm}, all curves merge into one at $T <
T_K$, showing the expected universal Kondo scaling behavior; see
\Fig{figS4}(b). Whereas the curves become separated at $T > T_K$,
indicating the systems are outside the Kondo regime.
It is important to point out that the second term in the exponent of
\Eq{eq:TK_symm} is unnegligible to achieve the universal scaling
behavior quantitatively.
Moreover, note that as the value of $U$ approaches $\Delta$, $T_K$ of
\Eq{eq:TK_symm} is no longer the true Kondo energy scale. This is
clearly manifested by the $U = 1.2\Delta$ curve in \Fig{figS4}(b),
where the $dI/dV$ versus $T/T_K$ data deviate significantly from the
universal scaling curve. Instead, it is found that by taking $U$ as the
Kondo energy scale, the universal scaling is retrieved; see the pink
squares in \Fig{figS4}(b). It is thus inferred that the Kondo energy
scale depends sensitively on the relative magnitudes of relevant
parameters ($U$, $\Delta$, and $W$).
For lower temperatures, a higher truncation level $L$ is necessary to
achieve quantitative convergence, which is computationally too
demanding for present HEOM approach.


By varying the value of $\ep_d$ or $U$ while fixing the others, the
SIAM is tuned away from the particle-hole symmetry. The degree of
asymmetry can be characterized by a parameter $\delta = \ep_d +
\frac{U}{2}$. It is well known that for an asymmetric SIAM
(corresponding to $\delta \neq 0$), its Kondo temperature deviates from
the form of \Eq{eq:TK_symm}.\cite{Zit06045312}
In particular, it had been derived by Krishna-murthy, Wilkins, and
Wilson\cite{Kri801044} that the Kondo temperature of an asymmetric SIAM
in the local moment regime ($T \ll -\ep_d \ll U$ and $\Delta \ll U \ll
W$) can be evaluated via
\begin{align}
   \ep_d^\ast &= \ep_d - \frac{\Delta}{\pi} \ln\left(\frac{-U}{\ep_d^\ast}\right), \nl
   \rho_0 J_K &= \frac{2\Delta}{\pi}\,
   \left( \frac{1}{|\delta - U/2|} + \frac{1}{|\delta + U/2|} \right), \nl
   \rho_0 K &= \frac{\Delta}{2\pi}\,
   \left( \frac{1}{|\delta - U/2|} - \frac{1}{|\delta + U/2|} \right), \nl
   \rho_0 \tilde{J}_K &= \rho_0 J_K \left[ 1 + (\pi \rho_0 K)^2\right]^{-1}, \nl
   T_K' & = 0.182\,|\ep_d^\ast|\,\sqrt{\rho_0 \tilde{J}_K}\,
   \exp\left(\frac{-1}{\rho_0 \tilde{J}_K}\right). \label{eq:TK_asym}
\end{align}
Here, $\ep_d^\ast$ is the renormalized on-site energy.
We examine a series of SIAMs in this regime, with fixed $\ep_d$ but
different values of $U$ (and hence different $\delta$).
The corresponding $dI/dV$ curves versus $T$ and $T/T_K'$ are depicted
in \Fig{figS5}(a) and (b), respectively.
The scattered curves in \Fig{figS5}(a) merge into one in
\Fig{figS5}(b), indicating that $T_K'$ of \Eq{eq:TK_asym} indeed
provides a universal Kondo energy scale for asymmetric SIAM systems.
Note that in \Fig{figS5}(b) the $U=1.6\,$meV data (represented by pink
triangles) deviate slightly from the universal scaling relation at low
temperatures, which is possibly due to the fact that such a value of
$U$ approaches to the reservoir bandwidth of $W = 2.0\,$meV.

\subsubsection{Spectral tail for Kondo resonance}
\label{sub2sec:ktail}

It has been predicted that for symmetric SIAM systems, the Kondo peak
of scaled spectral function has a slowly varying logarithmic tail as
follows,\cite{Dic014505}
\be
   A(\w) = \frac{1}{2\pi\Delta} \left\{
   \frac{1}{\left[(4/\pi)\,{\rm ln}(|\w'|) \right]^2 + 1} +
   \frac{5}{\left[(4/\pi)\,{\rm ln}(|\w'|) \right]^2 + 25}
   \right\}. \label{Aw-tail}
\ee
Here, $\w' = \w/\w_m' = \w / [\lambda e^{-\pi U/8\Delta}]$ with
$\lambda = {\rm min}(W, U/2)$ being the high-energy cut-off. It has
been verified that \Eq{Aw-tail} fits well with the NRG result at $T =
0$.\cite{Dic014505}
%

Figure~3(b) of main text plots the spectral function around the Kondo
peak calculated by the HEOM approach, which is compared with the
analytic curve of \Eq{Aw-tail}. Clearly, at relatively large scaled
frequencies, (say, at $\w/\w_k > 150$) the numerical results agree
remarkably with the analytic curve under all temperatures studied;
while in the low-frequency region, the HEOM results converge
consistently to the analytic curve (corresponding to $T = 0$) as the
temperature decreases. At even larger frequencies ($\w/\w_k > 500$),
the Hubbard peak starts to rise, which overwhelms the tail of Kondo
resonance peak, and hence this part of $A(\w)$ does not show up in the
comparison.

Figure~\ref{figS6} depicts the comparison between the HEOM calculated
$A(\w)$ for symmetric SIAMs with various values of $U$ and the analytic
expression of \Eq{Aw-tail}. Apparently, the numerical data agree better
with the analytic formula as $U$ increases. This is due to the fact
that the exponent $e^{-\pi U/8\Delta}$ involved in \Eq{Aw-tail} is
asymptotically exact in the strong coupling limit.\cite{Dic014505}

The above comparisons clearly affirm that the HEOM approach correctly
reproduces the asymptotic logarithmic scaling of Kondo resonance tail.

\subsection{Application to two-impurity Anderson model} \label{TIAM}

\begin{figure}
\includegraphics[width=0.8\columnwidth]{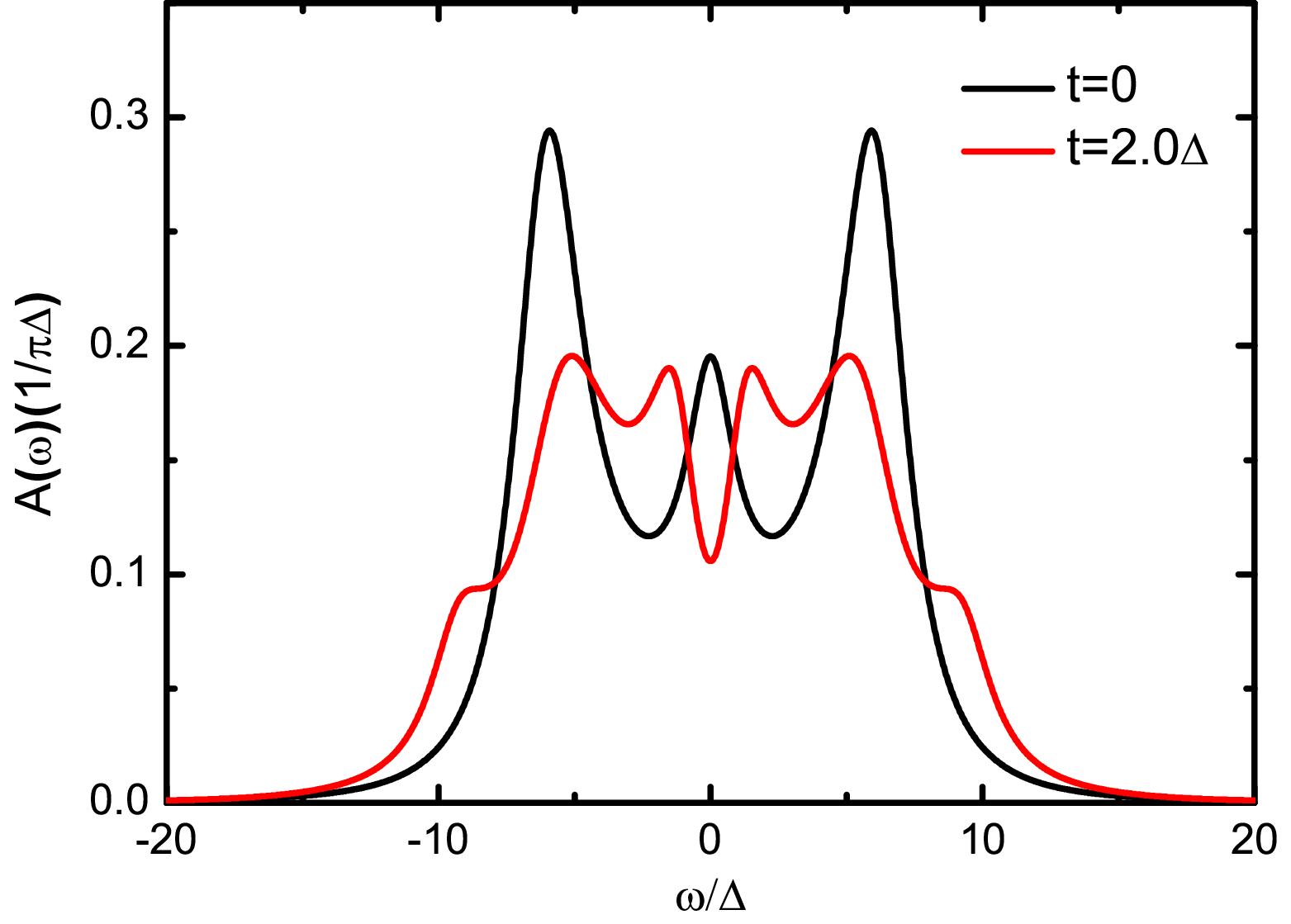}
\caption{The full spectral function of a symmetric TIAM
calculated by the HEOM approach at $L = 3$.
The same set of parameters as that adopted
for plotting Fig.\,3 in the main text are
considered here (in unit of $\Delta$):
$\ep_1 = \ep_2 = -5$, $U_1 = U_2 = 10$,
$W = 10$, and $T = 0.5$.} \label{figS7}
\end{figure}

\begin{figure}
\includegraphics[width=0.8\columnwidth]{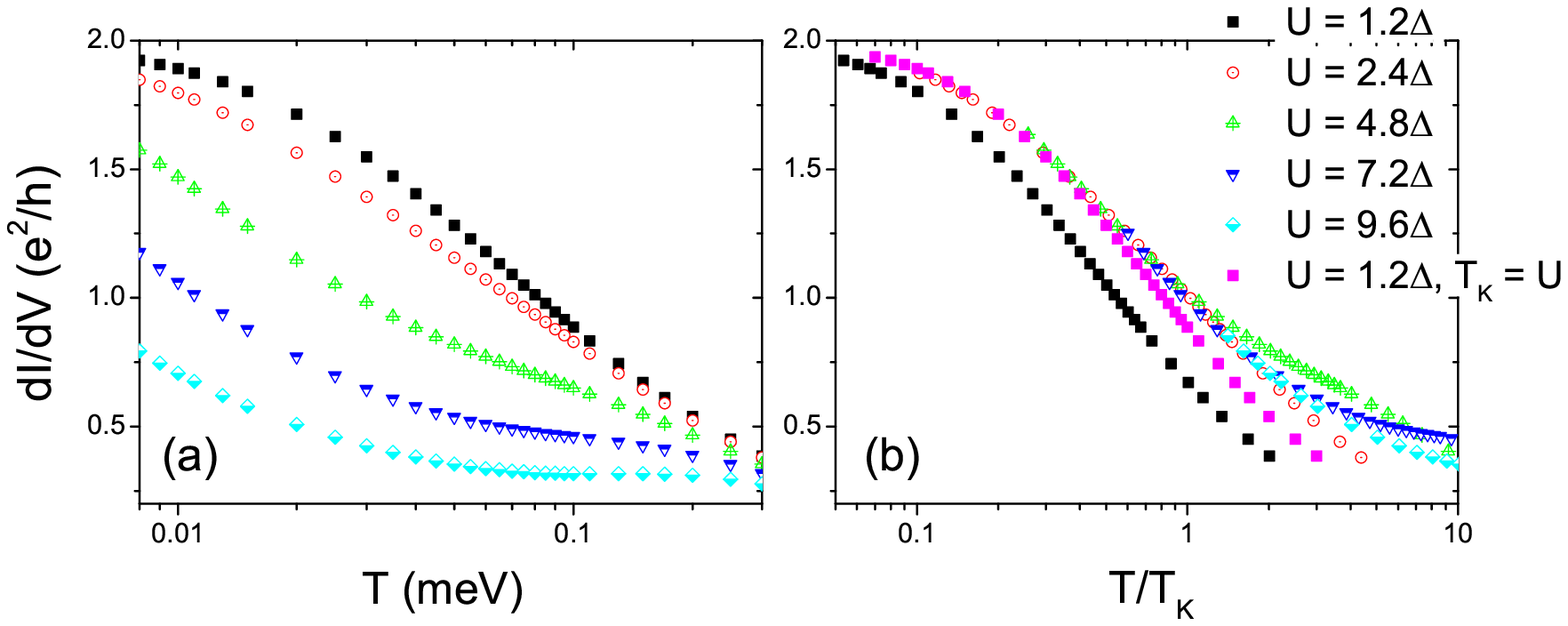}
\caption{
Zero-bias differential conductance (in unit of $e^2/h$) versus
(a) unscaled and (b) $T_K$--scaled temperatures,
for symmetric SIAM systems of different values of $U = -2\ep_d$.
The Kondo temperature $T_K$ is evaluated with \Eq{eq:TK_symm}, except
for the pink squares in (b) where $U$ is taken as the Kondo energy scale
and used to scale the temperature.
Other parameters are (in unit of meV): $W = 2$ and $\Delta = 0.0833$.
Calculations are done with the HEOM approach with $L = 4$.
} \label{figS4}
\end{figure}

\begin{figure}
\includegraphics[width=0.8\columnwidth]{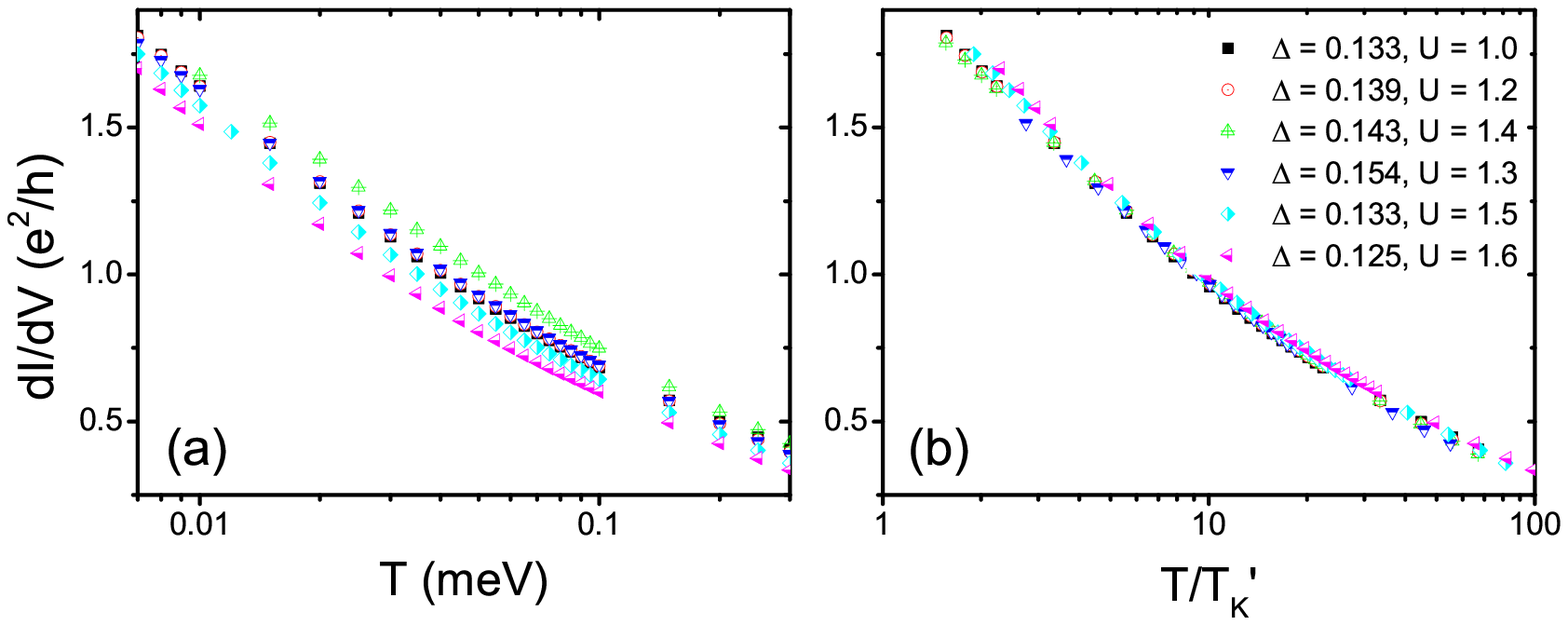}
\caption{
Zero-bias differential conductance (in unit of $e^2/h$) versus
(a) unscaled and (b) $T_K'$--scaled temperatures,
for asymmetric SIAMs of different $U$ and $\Delta$ in unit of meV.
The Kondo temperature $T_K'$ is calculated with \Eq{eq:TK_asym}.
Other parameters are (in unit of meV): $W = 2$, and $\ep_d = -0.2$.
Calculations are done with the HEOM approach with $L = 4$.
} \label{figS5}
\end{figure}

\begin{figure}
\includegraphics[width=0.8\columnwidth]{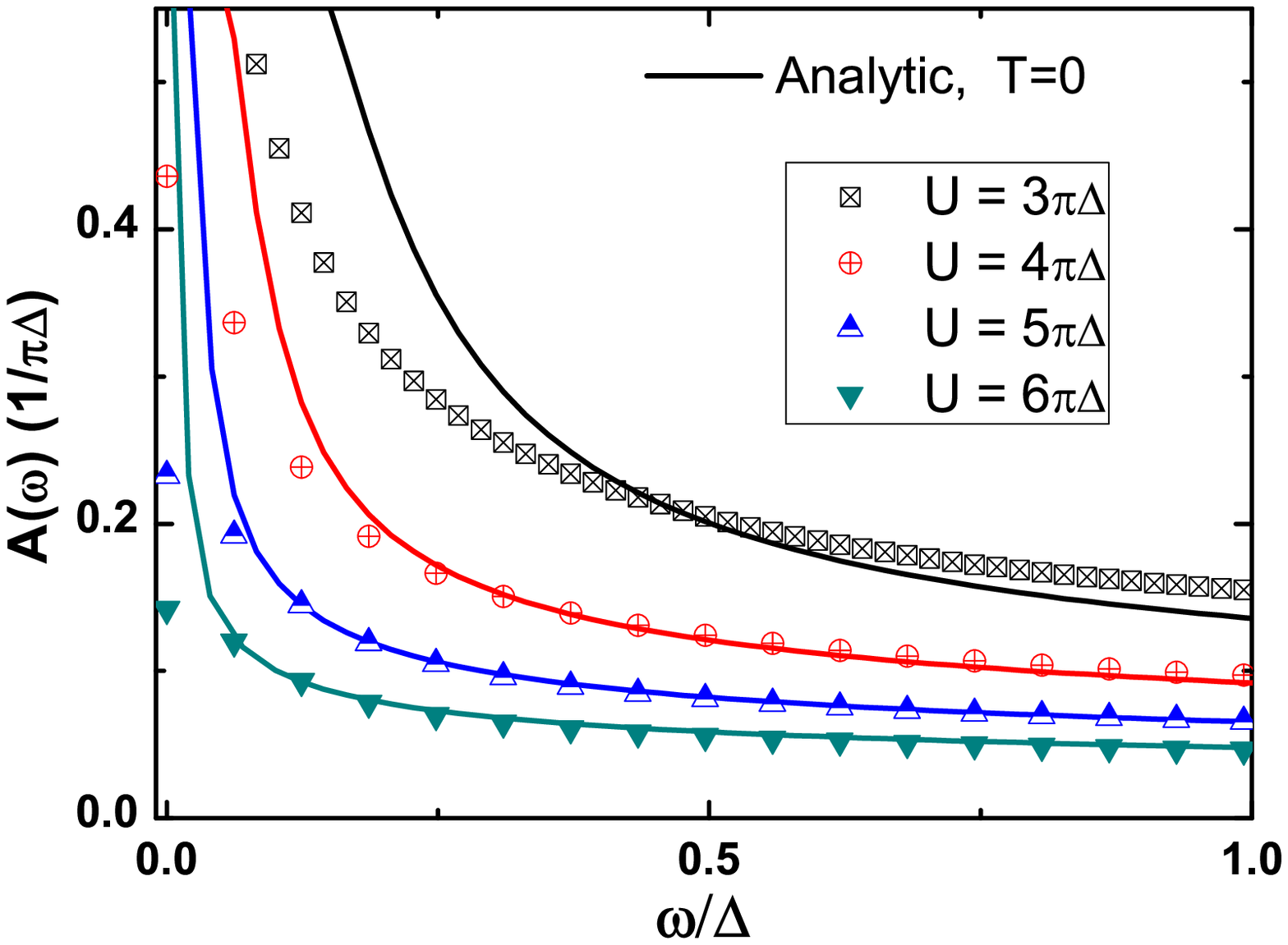}
\caption{
Spectral functions of symmetric SIAMs of various values of $U$.
The scattered data are the HEOM numerical results, while the solid curves
follow the analytic expression of \Eq{Aw-tail}.
Other parameters adopted in the HEOM calculations are: $W = 50\Delta$
and $T = 0.05\Delta$.
} \label{figS6}
\end{figure}

As described in the main text, we have
applied the HEOM approach to
two-impurity Anderson model (TIAM) system.
For clarity, a symmetric parallel-coupled two-impurity system
is considered. For the TIAM, we focus on how the inter-impurity
coupling $t$ affects the system dynamical properties
(\emph{e.g.} the spectral function of the impurities)
in the presence of strong \emph{e-e} interaction.
Fig.\,3 of the main text reveals that
with the increasing $t$,
the system exhibits a continuous crossover transition
from the Kondo singlet state of individual impurity
to the singlet spin-state between the two impurities.
This is affirmed by the impressive agreement between the
essential features of calculated $A(\w)$ and differential
conductance ($dI/dV$).
Actually, with our universal HEOM approach, the full spectrum of
$A(\w)$ can be obtained for any coupling strength $t$. As an example,
\Fig{figS7} depicts the full spectral functions of a symmetric TIAM at
$t=0$ and $t = 2.0\,\Delta$, calculated by the HEOM approach at $L =
3$. The same set of parameters as that to plot Fig.\,3 in the main text
are adopted (in unit of $\Delta$): $\ep_1 = \ep_2 = -5$, $U_1 = U_2 =
10$, $W = 10$, and $T = 0.5$. The sum rule $\int_{-\infty}^{\infty}
A(\omega)\, d\omega=1$ is satisfied to numerical precision.
One can see that the full spectrum resolves more resonance peaks at
high frequency range, which provides rich information
about the evolution of the energetic structure of
the impurities system with the variation in $t$.

In \Fig{figS7}, the three-peak-structure (the black line) at $t=0$ is
nothing but the duplication of the result for individual impurity, with
the itinerant electrons screening the local spin moment at each
impurity separately. At $t = 2.0\,\Delta$ (the red line), the
anti-ferromagnetic coupling ($J = 4t^2/U$) between the local spin
moments splits the Kondo peak at $\omega=0$. Meanwhile, the
inter-impurity coupling $t$ splits the high-frequency peaks at
$\omega=\varepsilon_{\mathrm{d}}$ and $U+\varepsilon_{\mathrm{d}}$.
Then, the three-peak structure of $A(\w)$ changes to a six-peak one, as
shown in \Fig{figS7}. The details will be systematically studied in our
future work.

\subsection{Application to multi-impurity Anderson models} \label{multi-IAM}

\begin{figure}
\includegraphics[width=0.8\columnwidth]{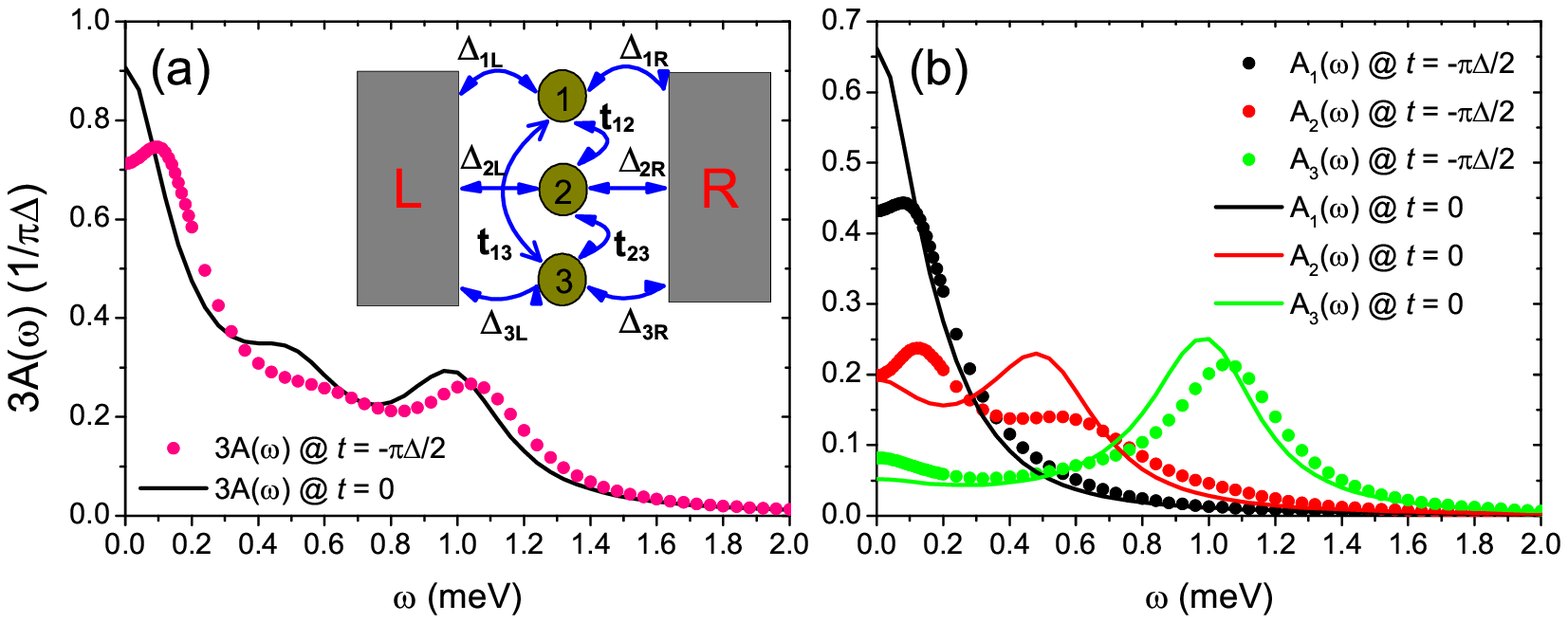}
\caption{
Spectral functions of a three-impurity Anderson model system
sketched in the inset of (a).
Panels (a) and (b) depict the spectral function of total system
$3A(\w) = \sum_i A_i(\w)$ and that of $i$th impurity $A_i(\w)$,
respectively.
The inter-impurity coupling strengths are set to $t_{12} = t_{23} = t$
and $t_{13} = 0$, \emph{i.e.}, only the nearest neighboring impurities
are considered to be coupled.
The lines and circles correspond to $t = 0$ and $t = -\pi\Delta/2$,
respectively.
Other system parameters are: $U_1 = -2\ep_1 = \pi\Delta$,
$U_2 = -2\ep_2 = 3\pi\Delta$, $U_3 = -2\ep_3 = 6\pi\Delta$,
$T = 0.75\Delta$, $W = 50\Delta$, and $\Delta = 0.1\,$meV.
Calculations are done with the HEOM approach with $L = 3$.
} \label{figS8}
\end{figure}

The HEOM approach is applicable to a general quantum impurity system,
and the numerical procedures are easily extended to more complex
systems than the single- and two-impurity Anderson models as discussed
in \Sec{SIAM} and \Sec{TIAM}.
To exemplify the applicability of the HEOM approach, we also perform
calculations on a three-impurity Anderson model (3IAM) with the system
Hamiltonian as follows,
\be
  H_{\rm sys} = H_1 + H_2 + H_3 + V_{12} + V_{13} + V_{23}, \label{hsys_3dot}
\ee
where $H_i = \sum_\sigma \ep_{i\sigma} \,
\hat{a}^\dag_{i\uparrow}\hat{a}_{i\uparrow} + U_i\,
\hat{n}_{i\uparrow}\hat{n}_{i\downarrow}$ and $V_{ij} = t_{ij}\,
(\hat{a}^\dag_{i\uparrow}\hat{a}_{j\uparrow} +
\hat{a}^\dag_{i\downarrow}\hat{a}_{j\downarrow}) + {\rm H.c.}$
Figure~\ref{figS8} exhibits the calculated system spectral functions of
3IAM for two cases: (i) $t_{12} = t_{13} = t_{23} = t = 0$, and (ii)
$t_{12} = t_{23} = t \neq 0$ and $t_{13} = 0$.
The variation of spectral density function induced by a small $t$ is
clearly demonstrated.
In particular, it has been verified that in the case of $t = 0$, the
calculated $A_i(\w)$ for $i$th impurity in a 3IAM is exactly identical
to $A(\w)$ of a SIAM with its Hamiltonian taking the form of $H_i$.
Note that \Eq{hsys_3dot} is in a general form, and does not distinguish
a degenerate case from a nondegenerate one. This infers that an
impurity with high degeneracy (such as a transition metal atom with
multiple $d$-electrons) is treated normally by the HEOM approach,
without additional computational effort.

The computational cost of HEOM approach increases exponentially with
the system size. The reason is mainly two-fold: (i) The system Hilbert
space (spanned by Fock states) expands as $4^N$, with $N$ being the
number of impurities. As a result, the system density matrix and each
ADO in the hierarchy is a $4^N$--by--$4^N$ matrix.
(ii) The number of distinct exponential functions used to resolve the
reservoir memory is $K = 2M N_\mu$, with $N_\mu$ being the number of
impurities that are coupled to the reservoirs.
Actually, in the present implementation of HEOM method, the major
bottleneck is the physical memory required to store all the ADOs
(rather than CPU time).

Fortunately, it is possible to improve substantially the efficiency of
HEOM based on physical considerations and making use of the sparse
nature of HEOM. This may extend further the applicability of HEOM to
more complex quantum impurity systems.
First, the $4^N$ growth of system Fock space may be suppressed by
imposing some physically reasonable constraint. For instance, with a
rather large Coulomb repulsion strength $(U_i)$, the double-occupancy
Fock states may be omitted, which reduces the scaling to $3^N$.
Moreover, quite often some off-diagonal blocks of density matrix and
ADOs are exactly zero, due to absence of quantum coherence between two
Fock states. These blocks may be safely removed, and hence lead to
saving of physical memory.
For resolution of reservoir memory, an efficient filtering algorithm
has been proposed by Shi~{\it et al.} for bosonic bath
environment,\cite{Shi09084105} which performs a screening process to
discard a large number of unimportant ADOs with negligible values, and
hence reduces greatly the size of hierarchy while preserving the
quantitative accuracy. Such kind of algorithms may be adopted to treat
larger quantum impurity systems coupled to electron (fermionic)
reservoirs, particular for the 5-band systems of significant
experimental relevance.

\section{Comparison between HEOM and other methods for quantum impurity systems} \label{comparison}
 In this section, we compare HEOM method and other methods
 frequently used for solving the Anderson impurity model, including
numerical renormalization group (NRG), strong coupling continuous time
quantum Monte Carlo (CTQMC), Green's function equation of motion
(GFEOM), exact diagonalization (ED), slave-boson mean-field theory
(SBMFT), and non-crossing approximation (NCA). For calculations
discussed below, unless stated otherwise, the parameters for the
Anderson impurity model are $\Delta = 0.02\,W$, $U=3 \pi \Delta$,
$\epsilon_d = - U/2$, and temperature $T=0.2 \Delta$. Here $W = 1$ is
the energy unit. Note that in the HEOM calculation, the energy unit is
set as $W = 10$. It does not change the dimensionless results.

\subsection{Comparison with numerical renormalization group method}

The NRG method has been regarded as the most accurate method for
solving impurity problems.\cite{Wil75773} This is because it can
resolve exponentially small energy scales, thanks to the logarithmic
discretization and iterative diagonalization algorithm. In order to
make conclusive comparison with NRG and HEOM, we first need to produce
accurate NRG data for the spectral function $A(\omega)$.

Here we use the full density matrix-NRG (FDM-NRG)
method\cite{Wei07076402,Pet06245114} to produce $A(\omega)$ at a finite
temperature. In order to get the most accurate spectral function, we
combine FDM-NRG with the self-energy method of
R.~Bulla\cite{Bul988365,Bul99136} and the z-average
method.\cite{Yos909403,Oli9411986} For broadening the
$\delta$-functions in $A(\omega)$, we adopt the following scheme:
A.~Weichselbaum's log-Gaussian functions with width $B_g$ are used for
$|\omega| > \alpha T$;\cite{Wei07076402} while normal Gaussian
functions with width $B_{\rm zero} = \alpha T$ are used for $|\omega| <
\alpha T$. In our calculation, we use $N_z = 8$, $\alpha = 0.5$, and
$B_g=0.2$ which are found to minimize the broadening error.

Besides the broadening error, there are two sources of error in
$A(\omega)$ from NRG. (1) Discretization error, usually described by
how large the logarithmic discretization parameter $\Lambda$ is
($\Lambda
> 1$). This error disappears in the limit $\Lambda \rightarrow 1$. (2)
Truncation error, described by how large the number of kept states
$M_{s}$ is. This error disappears in the limit $M_s \rightarrow
\infty$. Therefore, the parameters that influences the accuracy of
spectral function are $\Lambda$ and $M_s$.

\begin{figure}
\includegraphics[width=0.8\columnwidth, angle=270]{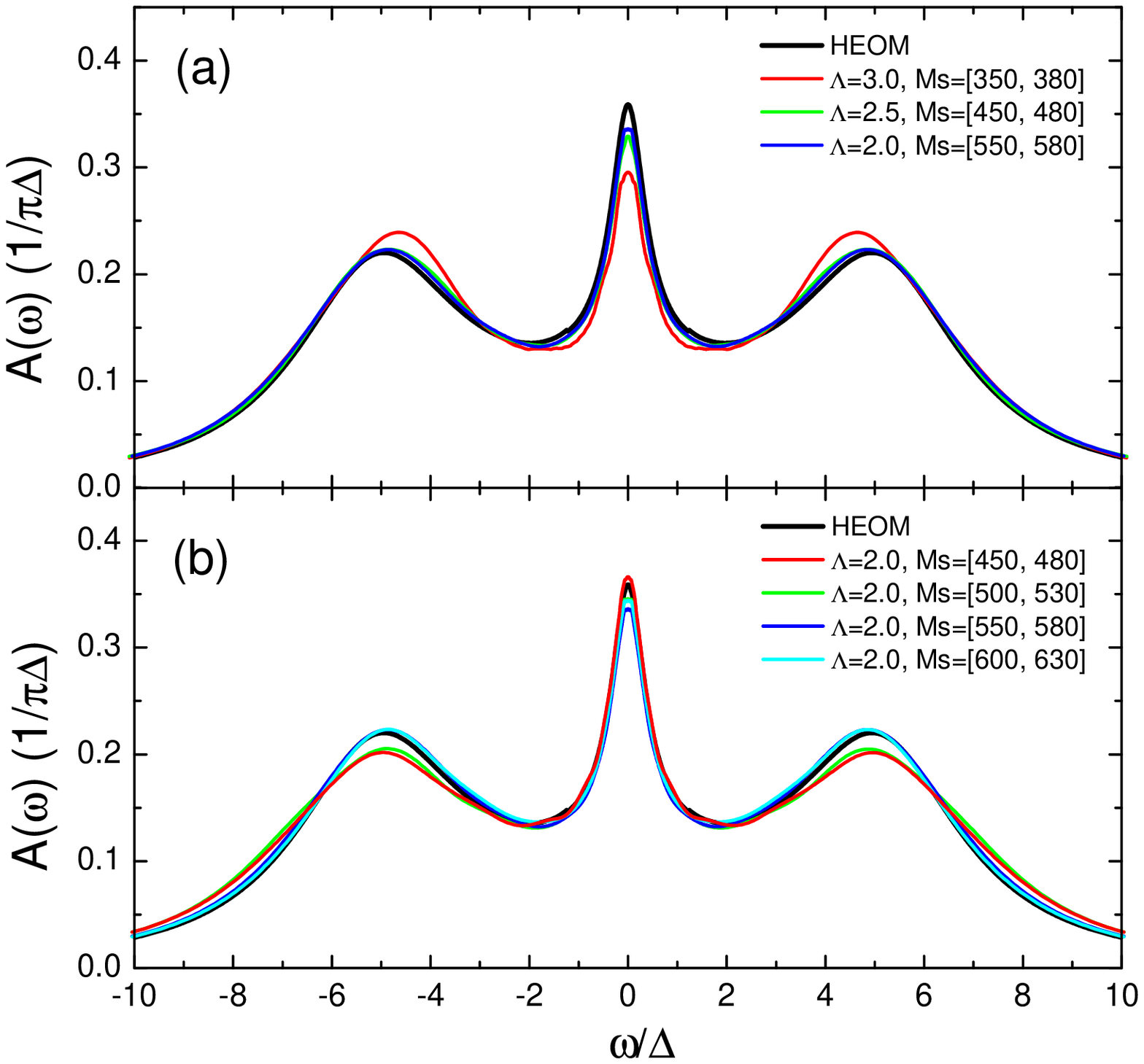}
\vspace{-2.0cm} \caption{(a) $A(\omega)$ from NRG using different
$\Lambda$ (red, green, and blue lines). For each $\Lambda$, we
choose a relatively large $M_s$. (b) $A(\omega)$ from NRG with
$\Lambda=2.0$ and different $M_s$ (red, green, blue, and cyan
lines). In both (a) and (b), HEOM curve is also shown for comparison
(black line). The parameters are $\Delta = 0.02\,W$, $U=3 \pi
\Delta$, $\epsilon_d = - U/2$, and temperature $T=0.2 \Delta$. $W$
is the energy unit.} \label{figS9}
\end{figure}

First, we examine the effect of decreasing $\Lambda$ to approach 1
where discretization error disappears. In \Fig{figS9}(a), we show
$A(\omega)$ for different $\Lambda$, with relatively large $M_s$ values
for each $\Lambda$. Since the larger $\Lambda$ is, the smaller $M_s$ is
required, we use different $M_s$ for different $\Lambda$. As $\Lambda$
decreases from $3.0$ to $2.0$, the weight in $A(\omega)$ transfers from
the Hubbard peaks to the Kondo peak, leading to an increase of the
height of Kondo peak and lower Hubbard peaks. As compared to HEOM
curve, the overall agreement with HEOM is improved when $\Lambda$
decreases.

Next, we examine the truncation error controlled by $M_s$. Note that
the curves in \Fig{figS9}(a) are not converged with respect to $M_s$.
Now we fix $\Lambda=2$ and increase $M_s$. It is seen from
\Fig{figS9}(b) that the curve for $M_s=[550, 580]$ are almost same as
that of $M_s=[600, 630]$ in large frequency regime. The height of Kondo
peak $A(0)$ converges as $M_s \rightarrow \infty$ in an oscillating
way. It begins to converge from below for the largest number of kept
states that we use, $M_s=[600, 630]$. We therefore believe that
$\Lambda=2.0, M_s=[600, 630]$ produces the best $A(\omega)$ within our
efforts.

It is seen in \Fig{figS9}(b) that the overall agreement between our
best NRG result and HEOM is very good. $A(0)$ has about $4\%$ relative
error. For larger frequencies the relative error is less than $4\%$.
Since the HEOM results are converged with respect to the level $L$
while our NRG curve has not converged to $\Lambda=1,
M_s\rightarrow\infty$, we therefore conclude that in Fig.\,1(b), the
accuracy of $A(\omega)$ from HEOM is close to or higher than the best
NRG result that we obtain. To reach such a comparable accuracy, the
calculation time of NRG and HEOM are also comparable, both are of the
order of a few hours on a workstation.

\subsection{Comparison with continuous time quantum Monte Carlo method}

CTQMC is in principle exact, provided that there is no minus sign
problem and the number of sampling tends to infinity.\cite{Gul11349}
Here we compare the spectral function from HEOM and the strong coupling
CTQMC calculations. For CTQMC, the imaginary time Green's function
$G(\tau)$ is first calculated. The spectral function $A(\omega)$ is
then produced by the maximum entropy method.\cite{ctqmc_code} The
transform of information from $G(\tau)$ to $A(\w)$ is an numerically
ill-posed problem. We find that the resulting spectral function depends
rather sensitively on the statistical distribution of the input
$G(\tau)$, which is controlled by parameters of CTQMC, such as nsweep
(number of sweeps), ntimes (number of discrete $\tau$'s), and iwmax
(number of Matsubara frequencies of input $\Delta(i\omega_n)$).

\begin{figure}
\includegraphics[width=0.8\columnwidth, angle=270]{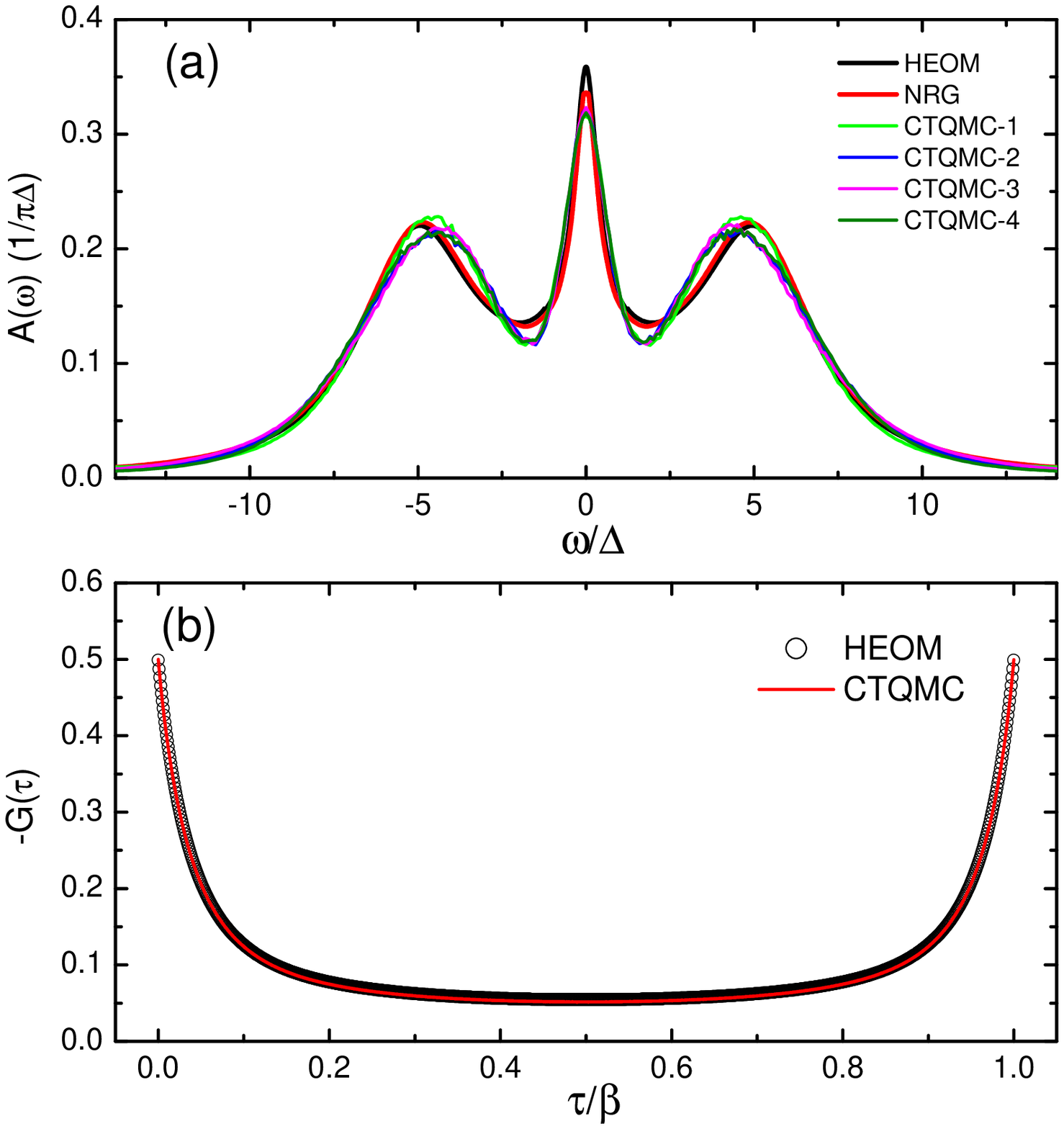}
\vspace{-1.0cm} \caption{(a) Comparison of spectral function from
HEOM (black line), NRG (red line), and the strong coupling CTQMC
(lines with other colors). The parameters for CTQMC-1 to CTQMC-4
are: (nsweep, ntimes, iwmax) = ($6\times10^{9}$, $2^{10}$, $2^{22}$)
, ($2.4\times10^{10}$, $2^{11}$, $2^{19}$), ($6\times10^{9}$,
$2^{13}$, $2^{22}$), and ($9.1\times10^{11}$, $2^{13}$, $2^{23}$),
respectivley. (b) Comparison of $G(\tau)$ bwteen HEOM (black circle)
and CTQMC (red line). The CTQMC data is obtained using the parameter
of the CTQMC-4 in (a). Other parameters are $\Delta = 0.02\,W$, $U=3
\pi \Delta$, $\epsilon_d = - U/2$, and temperature $T=0.2 \Delta$.
$W$ is the energy unit.} \label{figS10}
\end{figure}

This makes the error bars of $A(\omega)$ much larger than the
statistical error bars of $G(\tau)$. In \Fig{figS10}(a), we plot
$A(\omega)$ from CTQMC with different control parameters and compare
them with those from HEOM and NRG. It is seen that for these sets of
parameters, the variation of $\pi\Delta A(\omega)$ is $4 \times
10^{-2}$ around the Hubbard peaks, even though the statistical errors
of all the $G(\tau)$'s are already smaller than $6 \times 10^{-4}$.
Compared to HEOM result, the maximum relative error in $A(\omega)$ is
about $15\%$, mainly in the intermediate frequency regime. To elucidate
the source of discrepancy, we calculate $G(\tau)$ from $A(\omega)$ of
HEOM and compare it with the $G(\tau)$ directly from CTQMC. As shown in
\Fig{figS10}(b), they agree much better, with the maximum relative
error of $5\%$ at $\tau=\beta/2$.

Therefore, we conclude that although $G(\tau)$ agrees well between
CTQMC and HEOM, it is difficult to achieve quantitatively converged
spectral function $A(\omega)$ by CTQMC approach, due to the large
uncertainties associated with the analytical continuation of Green's
function to real energies with the maximum entropy method.

\subsection{ Comparison with refined Green's function equation of motion method}

Various equations of motion (EOM) methods using many-particle Green's
functions (GFs) or density operators as basic variables had been
developed.\cite{Ped05195330} However, very few method of this kind is
able to address strong correlation effects in quantum impurity system.
Luo~\emph{et al.} have proposed a refined GF--EOM
method,\cite{Luo999710} which yields $A(\w)$ of a SIAM in the Kondo
regime comparable to that by NRG method; see Fig.~2 of
Ref.~[\onlinecite{Luo999710}].
It is important to clarify that our HEOM approach is distinctly
different from the refined GF--EOM method of Luo~\emph{et al.} (and
many others) in terms of fundamental formalism, practical
implementation, and overall performance.

\begin{enumerate}
  \item \emph{Difference in fundamental formalism}

The GF--EOM method of Ref.~[\onlinecite{Luo999710}] uses many-body
GFs as basic variables, and the complete set of equations involves
explicitly all the single-electron states of the reservoirs (labeled
by $k$-index). It treats the system and reservoir degrees of freedom
on the same footing. This inevitably leads to a rather large number
of GFs to solve, even at a low level of hierarchy.
In contrast, our HEOM method uses reduced density matrix and
auxiliary density operators as basic variables. It is only the
reduced system (impurities) that is treated explicitly, and the
influence of electron reservoir of infinite degrees of freedom is
accounted for by statistical means. The only information required
for the reservoir is its memory content, which can be decomposed
efficiently into a number of characteristic dissipative modes.
Consequently, our HEOM method involves much fewer unknowns than the
GF--EOM, due to the statistical treatment for the electron
reservoir.

\item \emph{Difference in practical implementation}

Another key difference is that in our HEOM method the couplings
among different auxiliary density operators are determined
systematically from the Feynmann--Vernon path integral
formalism,\cite{Fey63118} and hence the HEOM can be cast into a
generic compact form; see Eq.~(1) of main text.
As a consequence, our HEOM method is applicable to more complex
quantum impurity systems such as multi-impurity and multi-reservoir
models without additional derivation effort.
In contrast, the extension of present GF--EOM method is much more
tricky, because the EOM of each individual GF requires analytic
derivation, and the complexity increases drastically as the model
becomes more complicated.

Furthermore, in our HEOM method the truncation of hierarchy is
carried out in a systematic way, so that it is easy to verify the
numerical convergence of resulting data with respect to truncation
level.
In contrast, the present GF--EOM method is hampered by lacking a
general recipe for truncation. In most of calculations, the
truncation scheme is strongly empirical and rather arbitrary.
Therefore, it cannot be regarded as a controlled approximate method.
In the work of Luo~\emph{et al.},\cite{Luo999710} a formal criterion
is introduced to guide the truncation, but still with the use of
empirical approximations to simplify the calculations.

\item \emph{Difference in overall performance}

As demonstrated in our manuscript, the spectral functions resulted
from our HEOM approach agrees quantitatively with those calculated
with the latest state-of-the-art NRG method. On the other hand, to
the best of knowledge, the most accurate GF--EOM method produces
only qualitatively correct results for the Anderson impurity model.
As shown in Fig.~2 of Ref.~[\onlinecite{Luo999710}], for
temperatures $T < T_{K}$, the relative deviation in $A(\w = 0)$ is
about $20\%$. In contrast, the HEOM results has a relative deviation
less than $4\%$ when compared with our best NRG result. Therefore,
it is apparent that HEOM generally outperforms the sophisticated
GF--EOM calculation in accuracy.

\end{enumerate}

\subsection{Comparison with exact diagonalization method }

The ED method is an important numerical method for solving many-body
problems with limited degrees of freedom.\cite{Dag94763} It is often
used as an impurity solver in the context of dynamical mean-field
theory.\cite{Caf941545, Si942761, Geo9613} As shown in Figs.\,1 and 2
in the main text, the HEOM approach can produce a smooth spectral
function similar to NRG by treating all reservoir degrees of freedom in
a statistical manner.
In contrast, the ED method uses a finite number ($N_b$) of
single-particle states to represent the electron reservoir.
Limited by computer memory, usually the adopted $N_b$ is not a very
large number, which might lead to strong quantum confinement effect,
due to the finite size of Hilbert space.
In practice, $N_b \leq 9$ is adopted for a full diagonalization
algorithm, and a somewhat larger $N_b \sim 15$ is feasible with the
Lanczos algorithm for ground-state properties at zero
temperature.\cite{Cap07245116,Lie12053201}

\begin{figure}
\includegraphics[width=0.8\columnwidth]{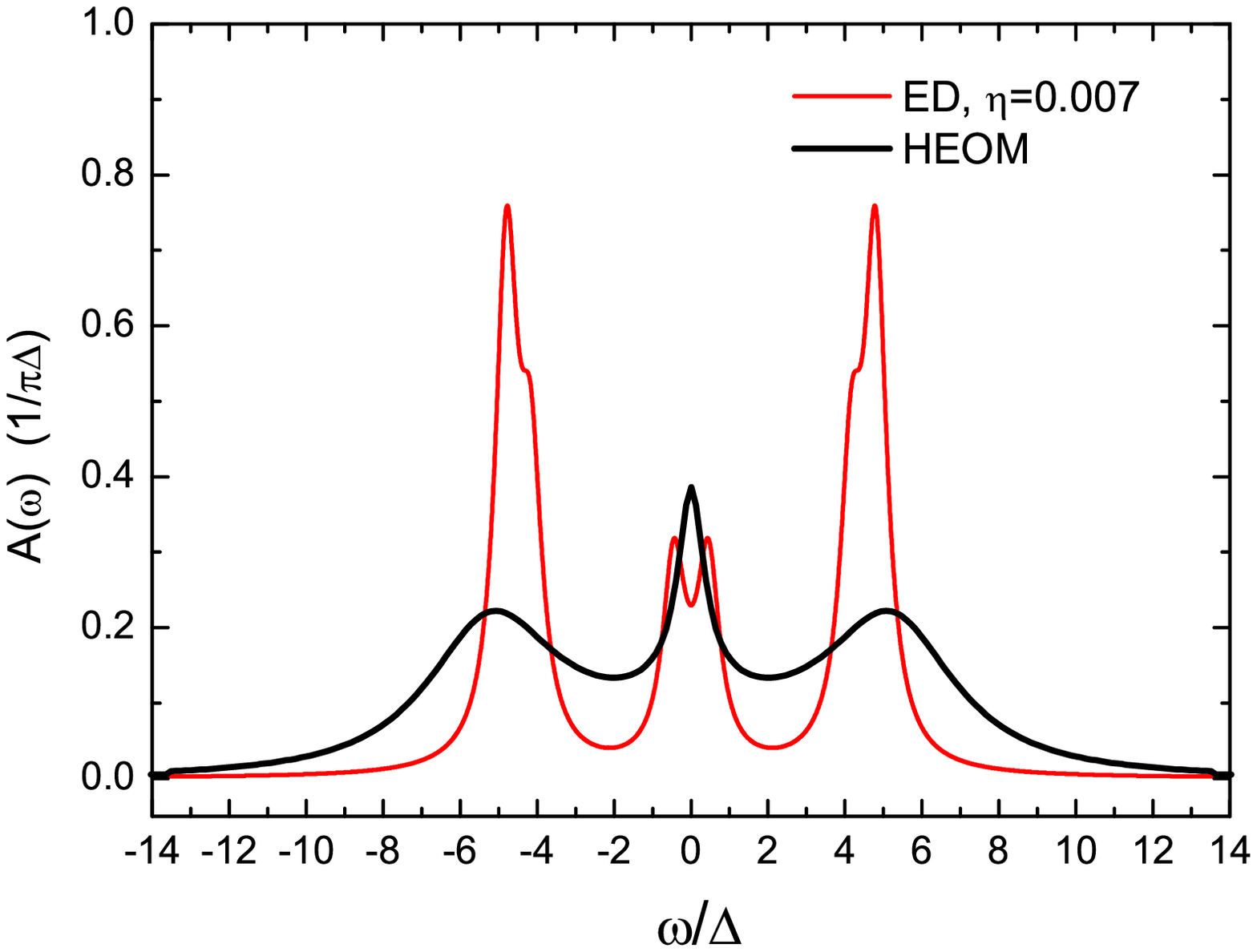}
\caption{Comparison of ED and HEOM spectral function. $A(\omega)$ from
ED ($N_b = 7$) is broadened with Lorentzian function of width
$\eta=0.007$. The parameters are $\Delta = 0.02 W$, $U=3.0\pi\Delta$,
$\epsilon_d = - U/2$, and temperature $T=0.2 \Delta$. $W$ is the energy
unit.} \label{figS11}
\end{figure}

Figure~\ref{figS11} shows the ED results for $U=3.0\pi\Delta$ using
$N_b=7$ bath sites, with $V_k$'s and $\epsilon_k$'s fitted from the
hybridization function $\Delta(i\omega_n)$. Here, an odd number of bath
sites is adopted to properly describe the noninteracting electron
reservoirs.\cite{Roz94535}
As shown in \Fig{figS11}, the Kondo peak and the Hubbard peaks are
resolved qualitatively by the ED method, but the Kondo peak is composed
of two small peaks with a splitting due to finite size effect. The
positions of the Hubbard peaks are close to $\pm U/2$, but their widths
are smaller than the HEOM results. This ED calculation takes about five
hours in our workstation, consumed mainly by the calculation of the
density of states. Such computational time is on the same order of
magnitude as the time of HEOM calculation shown in
Table~\ref{table:SIAM_highT} and \ref{table:SIAM_lowT}. Thus we
conclude that with similar computational cost, HEOM has much higher
energy resolution than ED.

\subsection{Comparison with slave-boson mean field theory and non-crossing approximation}

Both SBMFT and NCA are analytical approximation methods. In this
section, we compare their results with HEOM.

The Kotliar--Ruckenstein SBMFT\cite{Kot861362} is a convenient theory
to study the Anderson impurity model with finite $U$. It can
qualitatively describe features of the Fermi-liquid ground state, such
as the quasi-particle weight $z$ and the Kondo resonance peak in the
spectral function $A(\omega)$. The spectral function from SBMFT is
expressed as (for paramagnetic and particle-hole symmetric case),
 \begin{equation} \label{eq:1}
   A(\omega) = -\frac{1}{\pi} \text{Im} \frac{z}{\omega + i \eta - z \Gamma(\omega + i
   \eta)}.
\end{equation}
Here, $z$ is the quasi-particle weight, determined by a mean-field
equation. $\Gamma(\omega+ i\eta)$ is the Kramers--Kronig transformation
of the hybridization function $\Delta(\w)$. Therefore, $A(\omega)$ has
the form of non-interacting density of states, but renormalized by the
factor $z$.

\begin{figure}
\includegraphics[width=0.8\columnwidth, angle=270]{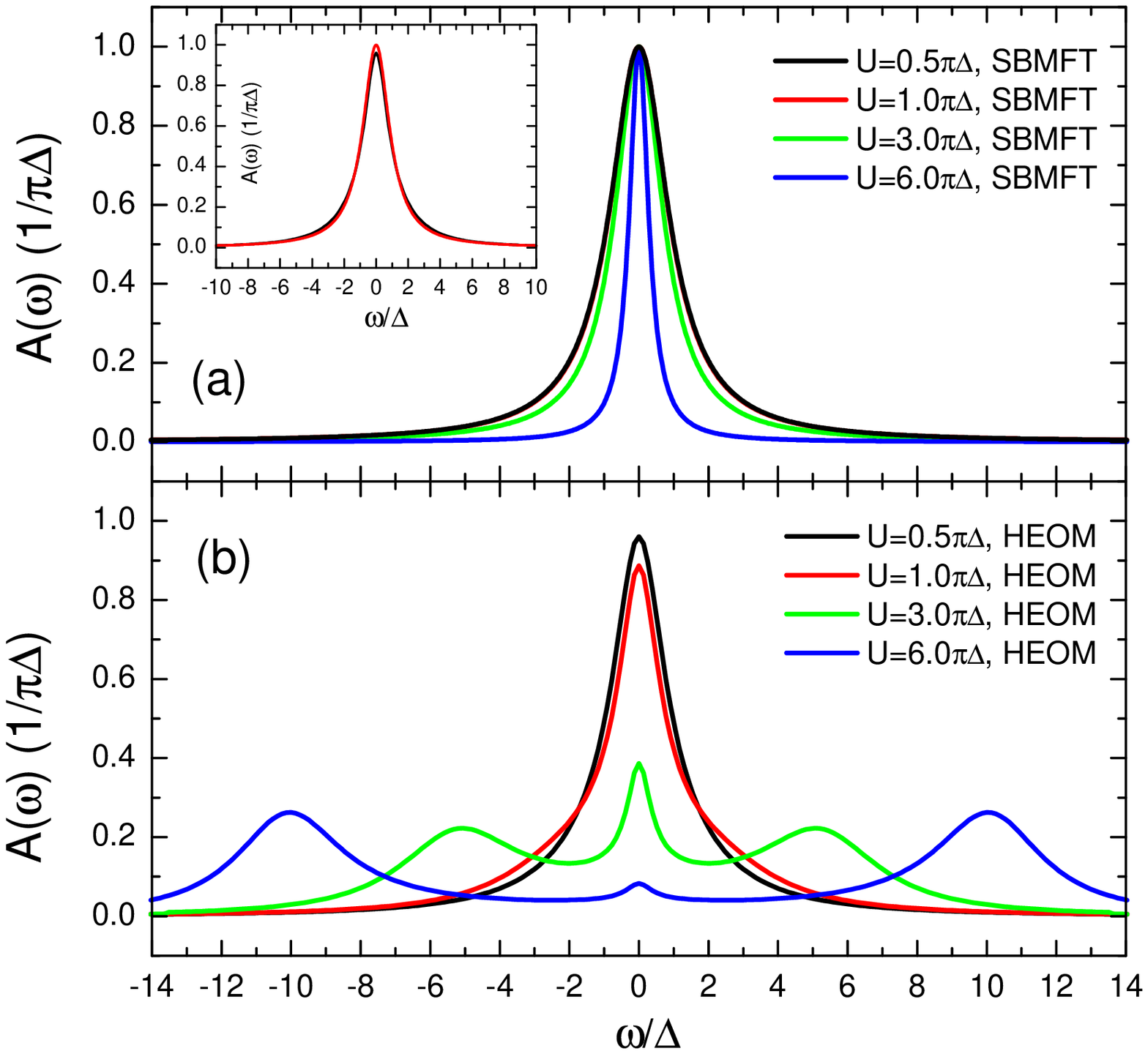}
\vspace{-1.0cm} \caption{(a) $A(\omega)$ from SBMFT for different
$U$'s. Inset: comparison of SBMFT (red line) and HEOM (black line)
result for $U=0.5\pi\Delta$. (b) $A(\omega)$ from HEOM for different
$U$. For (a) and (b), other parameters are $\Delta = 0.02\,W$,
$\epsilon_d = - U/2$, and temperature $T=0.2 \Delta$. $W$ is the
energy unit. } \label{figS12}
\end{figure}

In \Fig{figS12}, we compare $A(\omega)$ from SBMFT and HEOM at a finite
temperature $T=0.2\Delta$. In \Fig{figS12}(a), it is seen that the
Hubbard peaks at $\omega = \pm U/2$ are missing in $A(\omega)$ of
SBMFT. This is because SBMFT does not correctly describe the temporal
quantum fluctuations on the impurity site. The total weight of
$A(\omega)$ only contains the Fermi liquid quasi-particle weight $z$
which decrease from $1$ as $U$ increases. In the inset of
\Fig{figS12}(a), SBMFT and HEOM spectral function are compared for
$U=0.5\pi\Delta$. They agree quite well because for such a small $U$,
there is no local moment formation and the renormalized Fermi liquid
description is valid.

Another feature of $A(\omega)$ in \Fig{figS12}(a) is that the height of
Kondo peak $A(0)$ does not change with $U$. This is a feature of Fermi
liquid phase at $T=0$, but should not hold for the finite temperature
that we study here. This shows that for finite temperature, SBMFT
cannot correctly describe the suppression the Kondo peak by the thermal
fluctuation, while it is correctly described in the HEOM results shown
in \Fig{figS12}(b).

NCA is based on the partial summation of the diagrams in the
perturbative expansion of hybridization function under the slave-boson
representation. Complementary to SBMFT, NCA can handle the high energy
features such as the Hubbard peaks, but fails to describe the Fermi
liquid features in the low energy regime, such as the Kondo peak at
temperatures $T < T_{K}$ (Ref.~\onlinecite{Bic87845}). In contrast,
HEOM not only correctly produces the Hubbard peaks, but also produces
accurate Kondo peaks at the low temperature $T=0.2\Delta < T_{K}$, as
shown in the main text.

We conclude that both SBMFT and NCA work only in limited regime of
the parameter space, while HEOM applies to much wider area of the
parameter space, covering both low and high energies.

\clearpage

%
\begin{table}[h]
\centering
{ \small 
\begin{tabular}{r|r|r|r|r}
\hline \hline
$L$ & \# of unknowns & Memory (MB) & CPU time (s) & $n_d$~~~  \\
\hline
1  &        352 &    0.1 &    1.6  & \textbf{0.4}9156  \\
2  &      4,512 &    0.4 &    1.6  & \textbf{0.47}980  \\
3  &     29.152 &    2.0 &    2.3  & \textbf{0.476}02  \\
4  &    188,032 &   13.0 &    8.4  & \textbf{0.476}50  \\
5  &    868,096 &   59.7 &   39.7  & \textbf{0.47649}  \\
6  &  3,920,896 &  269.3 &  239.9  & \textbf{0.47649}  \\
\hline \hline
\end{tabular}
} 
\normalsize
\caption{Summary of HEOM calculations on
an SIAM with the parameters of (in unit of $\Delta$):
$T = 1.5$, $\ep_d = -5$, $U = 15$, $W = 10$,
and $\Delta_L = \Delta_R = 0.5$. The number of
occupied electrons on the impurity (per spin),
$n_d = {\rm tr}_{\s}(\hat{n}_\mu \rho_{\rm eq})$,
is calculated at various truncation tiers (up to $L = 6$).
The number of unknowns, computer memory, and CPU time are
listed for each $L$. The converged digits of the resulting
$n_d$ are highlighted with boldface at each $L$.
A total number of 4 Pad\'{e}
points are sufficient to
reproduce the reservoir correlation function quantitatively.
Calculations are done on a single PC with a 4-core
2.8\,GHz Intel(R) Core(TM) i5--2300 CPU.
}
\label{table:SIAM_highT}
\end{table}
%

%
\begin{table}[h]
\centering
{ \small 
\begin{tabular}{r|r|r|r|r}
\hline \hline
$L$ & \# of unknowns & Memory (MB) & CPU time (s) & $n_d$~~~  \\
\hline
1  &      1,056 &    0.1 &    1.6  & ~0.50000  \\
2  &     42,528 &    3.0 &    3.1  & ~\textbf{0.4}9163  \\
3  &    774,688 &   53.2 &   53.4  & ~\textbf{0.4}7832  \\
4  & 14,678,816 & 1008.0 & 1252.3  & ~\textbf{0.4822}4  \\
5  & 67,721,088 & 4650.1 & 9004.9  & ~\textbf{0.4822}6  \\
\hline \hline
\end{tabular}
} 
\normalsize
\caption{Summary of HEOM calculations on
an SIAM with the parameters of (in unit of $\Delta$):
$T = 0.075$, $\ep_d = -5$, $U = 15$, $W = 10$, and
$\Delta_L = \Delta_R = 0.5$. The number of
occupied electrons on the impurity (per spin),
$n_d = {\rm tr}_{\s}(\hat{n}_\mu \rho_{\rm eq})$,
is calculated at various truncation tiers (up to $L = 5$).
The number of unknowns, computer memory, and CPU time are
listed for each $L$.
A total number of 15 Pad\'{e} points are used to
reproduce the reservoir correlation function at all $L$,
except for $L = 5$ where 12 Pad\'{e} points are used.
Calculations are done on a single PC with a 4-core
2.8\,GHz Intel(R) Core(TM) i5--2300 CPU.
}
\label{table:SIAM_lowT}
\end{table}
%


%

\clearpage

%

\clearpage

%

\clearpage

%

\clearpage

%

\clearpage

%

\clearpage

%

\clearpage

%

\clearpage

%

\clearpage

%

\clearpage

%

\clearpage

